\documentclass{jfm}
\usepackage{graphicx}
\usepackage{epstopdf, epsfig}
\usepackage{amsmath,amssymb}
\usepackage{color}

\shorttitle{TRANSPORT EFFICIENCY OF METACHRONAL WAVES}
\shortauthor{S. Chateau, J. Favier, U. D'Ortona and S. Poncet}

\title{TRANSPORT EFFICIENCY OF METACHRONAL WAVES IN 3D CILIA ARRAYS IMMERSED IN A TWO-PHASE FLOW}

\author{S. Chateau\aff{1,2}
  \corresp{\email{sylvain.chateau@usherbrooke.ca}},
  J. Favier\aff{1},
  U. D'Ortona\aff{1}
 \and S. Poncet\aff{1,2}}

\affiliation{\aff{1}Aix Marseille Univ, CNRS, Centrale Marseille, M2P2, Marseille, France
\aff{2}Facult\'e de G\'enie, Universit\'e de Sherbrooke, Sherbrooke, Qu\'ebec, Canada}

\begin{document}

\maketitle

\begin{abstract}
The present work reports the formation and the characterization of antipleptic and symplectic metachronal waves in 3D cilia arrays immersed in a two-fluid environment, with a viscosity ratio of 20. A coupled lattice-Boltzmann - Immersed-Boundary solver is used. The periciliary layer is confined between the epithelial surface and the mucus. Its thickness is chosen such that the tips of the cilia can penetrate the mucus. A purely hydrodynamical feedback of the fluid is taken into account and a coupling parameter $\alpha$ is introduced allowing the tuning of both the direction of the wave propagation, and the strength of the fluid feedback. A comparative study of both antipleptic and symplectic waves, mapping a cilia inter-spacing ranging from 1.67 up to 5 cilia length, is performed by imposing the metachrony. Antipleptic waves are found to systematically outperform sympletic waves. They are shown to be more efficient for transporting and mixing the fluids, while spending less energy than symplectic, random, or synchronized motions. 
\end{abstract}

\section{Introduction}

Cilia and flagella are contractile hair-like structures, put into motion by biochemical energy, and protruding on the free surface of eukaryotic or prokaryotic cells. While prokaryotes are single-celled organisms, eukaryote cells have membrane-bound organelles and are found in every mammals. Many living organisms, going from the prokaryotic bacteria to mammals, use ciliary and/or flagellar propulsion as a swimming mechanism. Usually, flagella are external appendices used by micro-swimmers such as the algae \textit{Chlamydomonas reinhardtii} for locomotion purposes, while cilia are generally internal appendices, shorter and more numerous, used for moving materials such as nutrients, dusts, or proteins into living organisms. Ciliary propulsion is a universal phenomenon, and many examples could be cited. For the particular case of the human body, cilia are responsible for the left-right asymmetry of the heart in the early embryonic development, for the transport of nutrients in the brain, and for the transport of mucus in the mucociliary clearance process which is the background of the present work (see \citet{Satir} for a review about the structure and function of mammalian cilia).

During the breathing process, a large number of foreign particles (bacteria, dusts, pollutants or allergens) can penetrate the organism. The human body has then developed three mechanisms to protect itself from these particles: coughing, alveolar clearance and mucociliary clearance which occurs on the epithelial surface of the respiratory system.
In order to trap the particles, a layer of fluid called the Airways Surface Liquid (ASL) is covering the epithelial surface. Due to differences in the concentration of mucins inside the ASL, it is generally assumed to be the superposition of two different layers: the periciliary layer (PCL) and the mucus. The $7$ $\mu$m depth PCL is located between the epithelial surface from where the cilia protrude and the mucus phase just above it. Mainly composed of water and of a few mucins of low molecular weight, this phase, often considered as being a Newtonian fluid similar to water, is a kind of lubricant allowing the mucus to slip on it and the cilia to beat without too much viscous resistance. 

The mucus is a highly non-Newtonian fluid, with a strong visco-elastic behaviour. Additionally, the inner structure of the macromolecules composing it confers to the mucus a clear thixotropic behaviour, which is currently being studied and characterised to understand its inner rheological properties which exhibit a large variability \citep{Lafforgue}. 
Mucus is indeed composed of 95\% of water, but also contains macromolecules called the mucins \citep{Lai}.
It serves as a physical barrier against infectious agents and dusts, but also to humidify the air flowing into the respiratory system and to catch the particles. Its height varies between 5 to 100 $\mu$m depending on many factors including the position in the respiratory system \citep{Widi}, the pathology for a particular person, and the quality of the air inhaled. Its viscosity can also vary by several orders of magnitude within the same day \citep{Kirkham}.

Cilia are the other protagonist of the mucociliary process. They are organized as tufts (around 200 to 300 cilia per tuft) at the epithelium surface. Formed by 9 pairs of microtubules placed regularly in a circle and 1 pair of microtubule at the center (which form the so-called \textit{axoneme}), their purpose is to propel the surrounding fluid layers. Their diameter varies from 0.2 to 0.3 $\mu$m and their length from 6 to 7 $\mu$m \citep{Sleigh2}.

The cilia motion can be decomposed into two steps: a stroke phase and a recovery phase. The stroke phase is characterized by almost straight cilia orthogonal to the flow in order to maximize the pushing effect; while the cilia are bending during the recovery phase in a more inclined plane to get closer to the epithelial surface in order to minimize the viscous resistance, and therefore to reduce their impact on the flow. During the stroke phase, which takes around 1/3 of the total beating period, the tips of the cilia enter the mucus phase \citep{Widi}. Their beating frequency is estimated to vary between 10 and 20 Hz. Note that the spatial asymmetry is essential for the cilia to generate propulsion in creeping flows, while the temporal asymmetry (recovery phase longer than stroke phase) is not necessary to induce a mucus motion \citep{KhaderiBalt}.

It has been experimentally observed that cilia usually do not beat randomly \citep{Sleigh}, but adapt instead their beatings accordingly to their neighbours, giving birth to the so-called metachronal waves (MCW) observed at the tips of the cilia. MCW occur when adjacent cilia beat with a constant phase lag $\Delta \Phi$ between each other. For $0 < \Delta \Phi < \upi$, the MCW move in the opposite direction as the fluid propelled and are called antipleptic MCW. On the contrary, for $-\upi < \Delta \Phi < 0$, the MCW move in the same direction to the flow, and are called symplectic MCW. When the phase lag $\Delta \Phi$ is null, all cilia beat in a synchronized way. Finally, when the phase lag between neighbouring cilia is $\Delta \Phi=\pm \upi$, a standing wave appears.

The universality of ciliary propulsion has intrigued scientists for decades, and several studies were conducted in order to understand it. One actual objective is to be able to mimic this process in order to, for instance, create cilia-based actuators for mixing, use them as flow-regulator in microscopic biosensors, or as micropumps for drug-delivery systems \citep{Huaming, Chen}. Moreover, diseases such as asthma or Chronic Obstructive Pulmonary Disease (COPD) are related to the mucociliary clearance process \citep{Gardiner}. 
The objectives are to understand the underlying mechanism that allows hundreds of cilia to act as a whole for the transport of mucus and how it affects the flow generated. The results could bring a deeper understanding in such pulmonary diseases.

Numerically, Gueron and co-authors \citep{Gueron97,Gueron99} showed in the 90's that two neighbouring cilia beating randomly will quickly synchronize within a few beating cycle due to hydrodynamic interactions and form antipleptic MCW. This has been recently confirmed by \citet{Elgeti}, who observed the formation of MCW in a 3D single-phase environment. A certain degree of freedom in the beating pattern is required as shown in the theoretical work of \citet{Niedermayer}. They modelled the beating pattern of cilia by circular trajectories. By allowing some flexibility in the radii, they managed to introduce the coupling leading to MCW formation.
Among the different models used for the study of ciliary propulsion, the envelope model \citep{Taylor,Reynolds,Tuck,brennen,BlakeA,BlakeB} assumed that cilia are so densely packed that it is possible to consider their tips as an oscillating surface. Nevertheless, such configuration has only been observed in nature for symplectic metachrony. Moreover, this technique is limited to small amplitude oscillations and impose no slip and impermeability conditions at the oscillating surface. 
In the sublayer (or stokeslets) model \citep{Keller, Lighthill,Phan,BlakeC,Gueron97,Gueron99,Niedermayer,Gauger2009,Smith,Ding} the cilia are modelled by a distribution of stokeslets, which impose a force on the surrounding fluid. A proper mirror image of the stokeslets is required to impose the no-slip condition on the surface from where the cilia protrude. However, the presence of a wall is known to alter the nature of the far-field of the stokeslets, and it is thought that this can have important consequences on the hydromechanics of the cilia near the wall \citep{BlakeD}. Moreover, stokeslets can only be used for fluids with constant viscosities. Results using this method tend to show that symplectic metachrony would be more efficient for the mucus transport than synchronously beating cilia, and that antipleptic metachrony would induce the lower flow rate. The opposite was nevertheless obtained by \citet{Gauger2009} who found that antipleptic metachrony was more efficient than symplectic metachrony for a particular cilia spacing. For a distance between two cilia of $1.5L$, $L$ being the length of the cilia, they obtained an increase in the pumping performance of $40\%$ relative to a single cilium. However, while the beating pattern used in \citep{Gauger2009} was realistic, they considered a slow stroke phase and a quick recovery phase which is the opposite of what is observed in nature. Some authors \citep{Gueron97,Gueron99,KimNetz} showed that the energy spent by a cilium would decrease in the presence of metachronal motion.
Others tried to model the internal axoneme of a cilium. Among them, \citet{Mitran} used an overlapping fixed-moving grid formulation, coupling the finite volume and the finite element methods, to study the emergence of MCW. His model was very detailed and has lot of assets (two-layer flow, viscoelastic mucus) but required many experimental parameters and the impact of the MCW on the flow has not been considered so far.
\citet{Niedermayer} observed MCW formation in 1D cilia arrays. Those waves were stable only if the wavelength was 4 times higher than the cilia spacing. For the interested reader, a clear review of those computational modelings of the internal axoneme can be found in \citet{Fauci}. 
A different approach is to let the cilia adapt their motion in order to find the most energetically efficient ciliary beating pattern. In that context, \citet{Eloy} and \citet{Lauga} computed the shape and energy-optimal kinematics of cilia from an energetic point of view. They found that the optimal kinematics strongly depend on the cilium bending rigidity, and closely resemble the two-stroke ciliary beating pattern observed in natural cilia. Similarly, \citet{Osterman} computed the ciliary beating pattern with optimal pumping efficiency of isolated cilia and arrays.
Recently, new methods have been introduced such as the immersed boundary method (IB) used by \citet{Dauptain} to model the swimming of the pleurobrachia. \citet{Lukens} used an IB method to study the mixing produced by a carpet of cilia in the context of the mucociliary clearance process. A coupled immersed-boundary lattice Boltzmann method (LBM) was also used by \citet{Sedaghat} to study several parameters in a 2D configuration using an Oldroyd-B model for the mucus rheology. They found that the transport of mucus was maximized when considering the mucus as a Newtonian fluid.
While many studies regarding the emergence of MCW were conducted, only few addressed 3D configurations of the mucociliary clearance process. Among them, \citet{Elgeti} managed to observe symplectic and laeoplectic (perpendicular to the power-stroke direction) MCW formations. \citet{Ding} did not study the emergence of MCW but performed a comparative study of antipleptic metachrony versus symplectic metachrony in terms of transport efficiency and mixing. Their results showed that both antipleptic and symplectic MCW enhanced the fluid transport and the mixing, with the antipleptic waves being the most efficient ones. However, the two aforementioned studies only considered a single fluid layer.

In this work, by using the solver developed by \citet{JCPZHE} and already validated in similar configurations, the focus is placed on the emergence of MCW in a 3D two-phase flow configuration with a viscosity ratio of 20 and a large number of cilia. In particular, it is shown that a simple hydrodynamical feedback based on mechanical concepts can trigger the emergence of both symplectic and antipleptic MCW, while usually only a single layer of fluid is considered and only antipleptic MCW are seen to emerge. From a numerical point of view, the main advantage of the present method is its ease of implementation. Additionally, the local character of the collisions in the LBM allows an easy and straightforward parallelization of the code, making the simulation of a large number of cilia possible. The numerical method also possesses the following advantages: (i) viscosity ratios up to \textit{O}($10^{2}$) can be achieved \citep{Porter}, and (ii) the mucus-PCL interface emerges intrinsically from the model. To the knowledge of the authors, this solver is the only one that combines all these capabilities. The main contribution of this work is the thorough analysis of the advantages of the antipleptic and symplectic MCW over synchronized beatings by computing appropriate transport and efficiency ratios, mapping an inter-cilia spacing going from $1.67$ to $5$ cilia length. 
Finally, and for the first time, both the PCL and the mucus layer have been taken into account in the present study. The inherent advantages of the MCW for flow transport are studied by (i) considering the efficiency of the waves to transport the flow, (ii) comparing the flow generated and the volume transported by the different kinds of metachrony, (iii) comparing the energetic cost of the MCW, and (iv) analyzing the capacity of the waves to transport particles from an energetic point of view. 
The following analysis shows that antipleptic MCW are always the most efficient to transport mucus. 

The remainder of this manuscript is organized as follows. In Section \ref{sec:Num}, details about the numerical method are given, and the different quantities used are introduced. In Section \ref{sec:Result}, the results are presented starting from the emergence of MCW by considering the fluid feedback onto the cilia; and then a parametric study is done to quantify the impact of both the MCW and the cilia spacing on the transport of mucus. A summary of the results and the future perspectives conclude this manuscript in Section \ref{sec:Concl}.

\section{Numerical method}\label{sec:Num}

From the numerical point of view, modeling the complete problem under realistic conditions remains a huge challenge and several assumptions need to be made to simplify the problem, while keeping the essential ingredients of the physical mechanisms in the model. In particular, it is assumed that there is no mass transfer at the wall, that both fluids are Newtonian and that the cilia are equally spaced filament-like structures. More assumptions are detailed in this section, together with the numerical setup and geometry.

\subsection{Geometrical modeling and beating pattern}\label{ssec:geometrical}

\begin{figure}
\centering
\begin{tabular}{cc}
\includegraphics[scale=0.25]{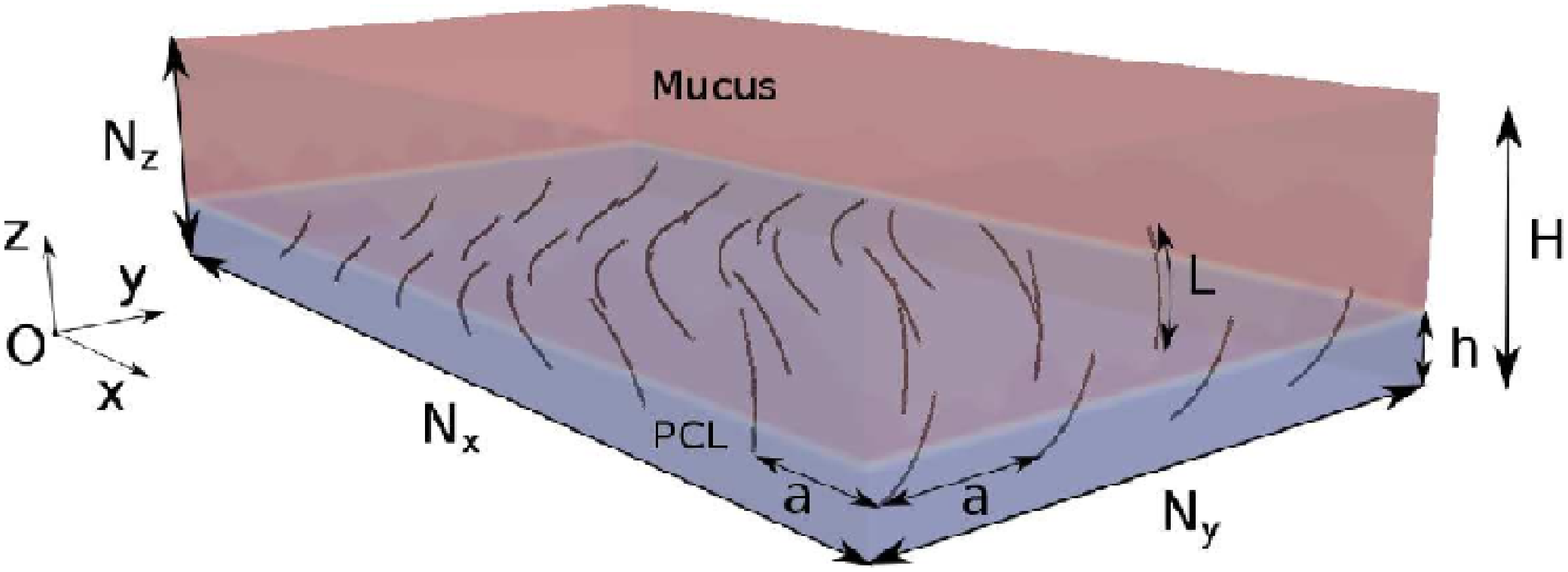} & \includegraphics[scale=0.18]{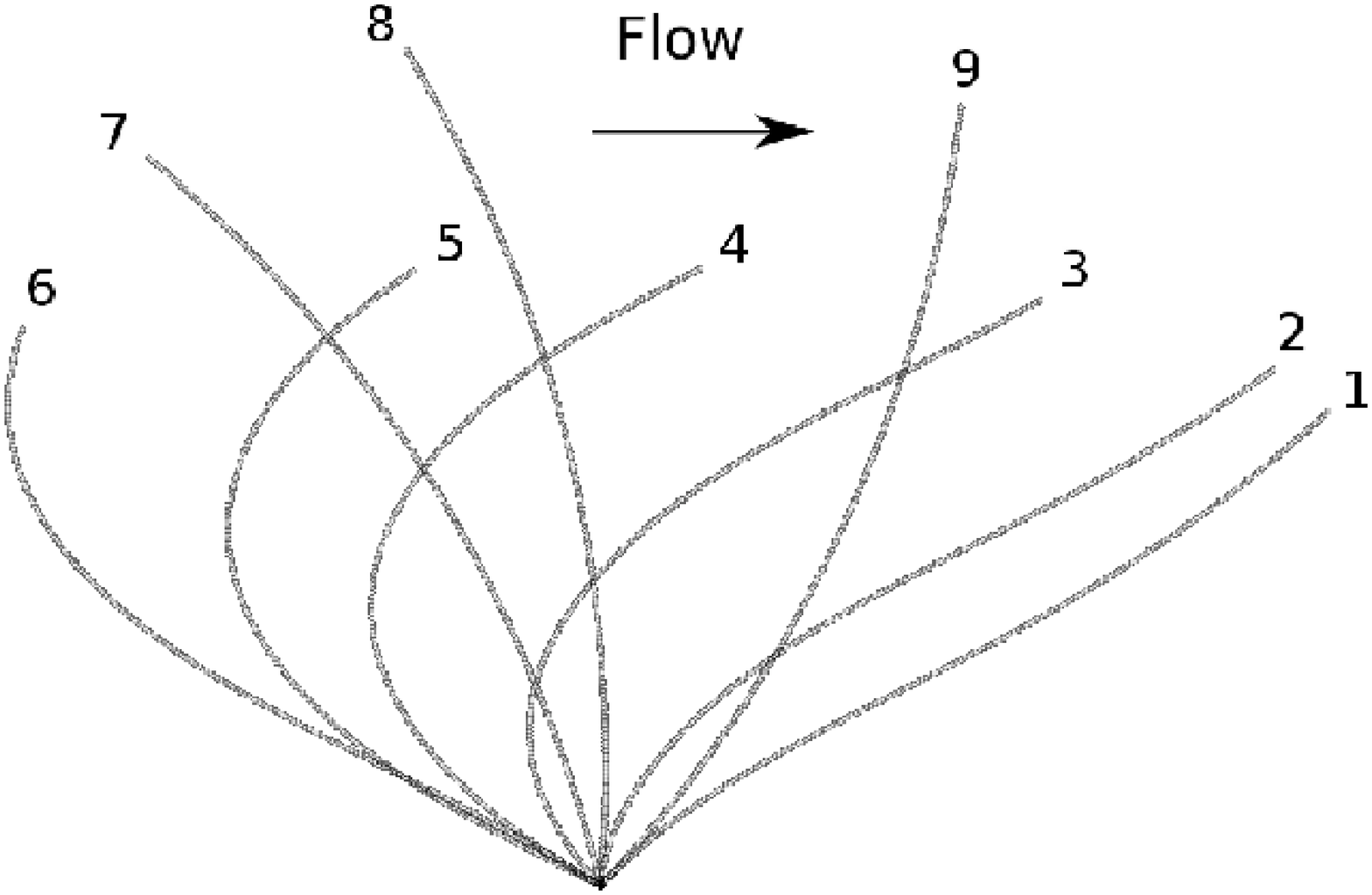} \\
(a)&(b)\\
\end{tabular}
\caption{(a) Schematic view of the computational domain. The present case corresponds to an antipleptic MCW. The domain is filled with PCL (in blue) and mucus (in red). (b) Beating pattern of a cilium with the parametric equation used. Steps 1 to 6 correspond to the recovery phase, and steps 7 to 9 to the stroke phase.}
\label{Figure1}
\end{figure}

The computational domain is a box with a regular mesh composed of $N_{x} \times N_{y} \times N_{z}$ points. The cilia are equally spaced along the bottom ($x$-$y$) wall, such that their base points are located at $z = 0$ (see figure \ref{Figure1}(a) for a schematic view of the geometry). Their motion is imposed to be in the $x$-direction only. The spacing between two adjacent cilia is denoted $a$ in the $x$ and $y$-directions. The wavelength $\lambda$ of the MCW is set such as $\lambda=N_{x}$ when the metachrony is forced.  
For the case of synchronously beating cilia, $N_{x}$ is chosen to be $8a$.
The length of the cilia $L$ is set to $15$ lattice units (\textit{lu}) 
except when stated differently; and $200$ snapshots in time per beating cycle are uniformly distributed to model their motion. The ratio $h/H$ between the PCL thickness and the height of the domain is fixed to $0.27$ for all simulations. 

The equations of motion for the cilia are inspired from \citet{Chatelin} and reproduce the beating pattern by resolving a 1D transport equation along a parametric curve. Let $P(\zeta,t)$ be the position of the curve at time $t$ and at a normalized distance $\zeta$ from the base point of a cilium. With appropriate boundary conditions, a realistic beating pattern can be obtained using the following transport equation:
\begin{align}
\frac{\partial P'}{\partial t} + \nu(t)\frac{\partial P'}{\partial \zeta} = 0 \quad \text{BC:}\left\{
 \begin{array}{ll}
 P(0,t)=(0, 0, 0) \\
 P'(0,t)=(2\cos(2\upi t /T),0,\cos(2\upi t /T))
 \end{array}
\right.
\label{1}
\end{align}
with $\nu(t)=[1+8 \cos^{2}(\upi(t+0.25T)/T)]/T$ being the viscosity of the surrounding fluid, $T$ the beating period, and $P'=\partial_{\zeta} P$.  
The resulting angular amplitude between the beginning and the end of a stroke phase is $\theta=\frac{2\upi}{3}$, which agrees well with experimental data \citep{Sleigh2}. Note that a 3D beating pattern would allow to achieve more realistic simulations while being more computationally expensive. Hence, the choice has been made to use this 2D beating pattern. It captures the essential ingredients of the beating, and as cilia have a diameter smaller than a lattice unit, the difference of induced flow between two cilia overlapping in 2D or slipping onto each other in 3D is very small.
Figure \ref{Figure1}(b) gives a view of the beating pattern obtained by resolving Eq. (\ref{1}).
Note that with the present model both phases take the same amount of time while moving in the same ($x$,$y$) plane. This choice has been made in order to only study spatially asymmetric motions, which are the only mechanisms effective at low Reynolds numbers. The temporal asymmetry is indeed a mechanism that can enhance the flow only when inertial forces are no longer negligible \citep{KhaderiBalt}.
Nevertheless, when the feedback of the fluid is taken into account (as in \S \ref{ssec:MCW}), a non symmetrical motion will develop, with a stroke phase slower than the recovery phase. More details regarding this temporal asymmetry are given in \S \ref{par:feedback}.

The PCL is set such that it fills the region going from the bottom of the domain ($z=0$) up to an altitude of $h$. In all the simulations, the value of $h=0.9L$ has been used in order to allow the tips of the cilia to emerge into the mucus layer during their stroke phase as observed in real epithelium configurations. 

Both PCL and mucus are considered to be Newtonian fluids. The kinematic viscosity of the mucus is $\nu_{m}=10^{-3}$ m$^{2}$/s, and the viscosity ratio $r_{\nu}=\nu_{m}/\nu_{PCL}$ between the mucus and PCL is set to 20. It has indeed been recently shown \citep{ChatelinPOncet} that mucus transport was maximized for viscosity ratios ranging from 10 to 20 with a stiff transition between the two fluid layers.
The beating period of the cilia is equal to $N_{it}\times dt$ (with $dt=1$ using the classical LBM normalization), $N_{it}$ being the number of iterations for a cilium to perform a complete beating cycle. 
An oscillatory Reynolds number $\Rey^{osc}$, based on the velocity of the cilia's tips $U_{cil}=\frac{2\theta L}{T_{osc}}$ can now be defined:
\begin{align}
\Rey^{osc}=\frac{ U_{cil}L}{\nu_{mucus}} =\frac{\frac{\frac{4\upi}{3}}{T_{osc}}L^{2}}{\nu_{mucus}} = \frac{\omega L^{2}}{\nu_{mucus}}
\end{align}
with $\omega$ being the angular frequency of the cilia beating. With realistic physical quantities corresponding to the ciliated epithelium surface, the value of $\Rey^{osc}$ is of the order of $10^{-5}$ (using $L\approx 10^{-5}$ m, $\nu_{mucus}\approx 10^{-3}$ m$^{2}$/s and $U_{cil}\approx 10^{-3}$ m/s). 

To avoid running simulations at such a low Reynolds number which would require a very high number of iterations using a lattice Boltzmann scheme, a higher Reynolds number was chosen: $\Rey=20$. Indeed, achieving simulations at $\Rey<1$ using a LB scheme, while still describing well the cilia ($L=15$ lu at least), requires a lot of CPU resources.
Thus, inertial effects are introduced in the model but it has been carefully checked that the results remain the same for creeping flows ($\Rey=\text{\textit{O}}(10^{-2})$) by  comparing them to the results obtained with LBM formulation designed for Stokes flows \citep{GuoZHAOLI, Zou}. The differences are found to be less than 7\% on the transported velocity for all phase lags $\Delta \Phi \neq 0$. The corresponding beating patterns of the cilia shown in the following are similar also in terms of vorticity generation and the same qualitative coordination concerning the antipleptic/symplectic behaviour is observed. However, for the particular case of synchronously beating cilia (i.e. $\Delta \Phi=0$), the inertial effects will play a non-negligible role in the flow dynamics. Despite being weak, they will indeed cancel the reversal of the flow that should occur when the cilia are in the recovery phase. More details regarding the inertial effects will be given in \S \ref{sssec:Reynolds}.

Finally, it is worth noticing that the lattice has been chosen such that the numerical diffuse diameter of the cilia due to IB method corresponds to real cilia diameter ($\approx 0.3 \mu$m). Hence, a realistic drag is taken into account in the present study.

\subsection{Algorithm}\label{ssec:Algo}

The numerical model is described in \citet{JCPZHE}, and validated on several configurations involving flexible and moving boundaries in multiphase flows, with a 2$^{nd}$ order accuracy. 
Briefly, the idea is to add a forcing term $F_{i}^{\sigma}=\boldsymbol{F}_{\sigma}^{SC}+\boldsymbol{F}_{\sigma}^{IB}$ to the discrete LB equation for each $\sigma$ fluid component. $\boldsymbol{F}_{\sigma}^{SC}$ is an interparticle potential force that takes into account the fluid-fluid cohesion forces \citep{ShanChen}, and $\boldsymbol{F}_{\sigma}^{IB}$ is an IB-related force for ensuring the no-slip condition at the fluid-solid interface.

The fluid part is first solved on a Cartesian grid with LBM using the Bhatnagar--Gross--Krook (BGK) operator and a D3Q19 scheme. The collision and streaming steps proper to the LBM method are first performed. The model of \citet{Porter} is used to model the two-phase flow as it allows to minimize the magnitude of spurious currents near the fluid-fluid interface. More importantly, it also allows to consider higher density or viscosity ratios.
Then, values for the fluid velocity are interpolated at the Lagrangian points. It allows to compute an IB force to be spread onto the neighbouring Eulerian fluid nodes in order to ensure the no-slip condition along the cilia. The macroscopic fluid velocity is then updated.
The motion of each cilium is decomposed into a finite number of steps (snapshots) during a period. If necessary, an interpolation can be done in order to have the velocity values along the cilia in between two steps. Note that the geometric shape of the beating is fixed in all simulations and is not impacted by the feedback law introduced in \S \ref{par:feedback}, which only affects the duration of the recovery and stroke phases.

Since the model of \citet{Porter} uses a Shan-Chen (SC) repulsive force \citep{Shan1993, ShanChen}, surface tension effects emerge intrinsically at the PCL-mucus interface. Hence, a sharp interface between the mucus and the PCL can be maintained at any time. However, small diffusion effects might occur on the lattices bordering the interface. Additionally, the cilia which enter the PCL-mucus interface may punctually induce a small mixing. This is however corrected by the SC force which ``unmixes'' the two fluids.

Periodic boundary conditions are used in the $x$ and $y$ directions, while no-slip and free-slip boundary conditions are used at the bottom and top walls respectively.
The size of the computational domain ranges from 50 \textit{lu} to 400 \textit{lu} depending on the configuration considered, except for the size along the $z$ direction which is always set to 50 \textit{lu}.
 
Taking advantage of the local character of the LBM algorithm, the code is parallelized using MPI libraries (Message Passing Interface), by splitting the full computational domain into $9$ sub-domains of size $(N_{x}/3,N_{y}/3,N_{z})$.
More details on the numerical model can be found in \citet{JCPZHE}.

\subsection{Feedback of the fluids onto the cilia}\label{par:feedback}
The basic idea is to modulate the beating motion of the cilia as a function of the fluids motion. To do so, it is assumed that all cilia follow the same beating pattern, meanwhile a feedback of the fluids, which consists in accelerating or slowing down the motion of the cilia, is introduced. 

Each cilium is discretized with $N_{s}=20$ Lagrangian points. Let $s$ be the subscript corresponding to the $s^{\text{th}}$ Lagrangian point, starting from the base tip at $s=1$, and $\boldsymbol{V}_{i}^{s}$ the velocity on the $s^{\text{th}}$ Lagrangian point of the $i^{\text{th}}$ cilium. For each cilium, we define the average velocity over all Lagrangian points $\boldsymbol{V}_{i}$, which is linked to the number of steps (snapshots) this cilium will skip during one iteration of the fluid solver. The fluid feedback onto the cilia thus consists in modifying the norm of the velocity vector $\| \boldsymbol{V}_{i}\|$, while its direction remains unchanged.

\begin{figure}
\centering
\includegraphics[scale=0.3]{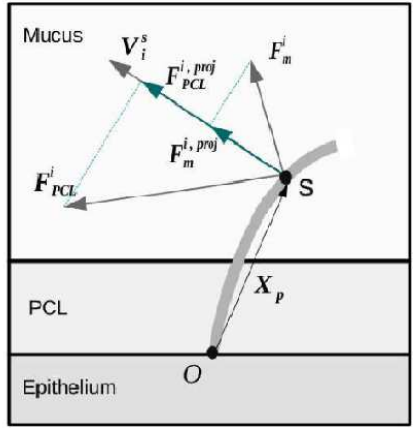}
\caption{Schematic view of a cilium with the corresponding forces exerted on the fluids.}
\label{Figure2}
\end{figure}

The feedback is computed in three steps. First, the IB forces corresponding to mucus and PCL are projected onto the corresponding velocity vectors for each Lagrangian point. Then, an estimate of the feedback is computed based on the torques of the forces for each Lagrangian point. Finally, the beating pattern of the cilia is adjusted at the beginning of the next time step. The different forces and geometrical variables used are illustrated in figure \ref{Figure2}, and the three steps explained below.

\begin{enumerate}

\item The interpolated IB forces applied by the $i^{\text{th}}$ cilium onto the fluids - respectively $\boldsymbol{F}_{m}^{i}$ for the force imposed on the mucus phase and $\boldsymbol{F}_{PCL}^{i}$ for the force imposed on the PCL phase-, are projected on $\boldsymbol{V}_{i}^{s}$:
\begin{align}
\boldsymbol{F}_{*}^{i,proj}=\frac{\boldsymbol{F}_{*}^{i}\boldsymbol{\cdot} \boldsymbol{V}_{i}^{s}}{\| \boldsymbol{V}_{i}^{s} \|^{2}}\boldsymbol{V}_{i}^{s},
\end{align}
where $*$ stands for ``$m$'' or ``PCL'' depending on the position of the Lagrangian node.
In order to take into account the difference of viscosity between the two layers, the forces $\boldsymbol{F}_{m}^{i,proj}$ and $\boldsymbol{F}_{PCL}^{i,proj}$ are weighted by a term under 
the form: $\frac{\tau_{*}}{\tau_{m}+\tau_{PCL}}$. The total projected force of the fluids onto the $s^{\text{th}}$ segments of the $i^{\text{th}}$ cilium $\boldsymbol{F}_{fluids\rightarrow cilia}^{i}$ writes:

\begin{align}
\boldsymbol{F}_{fluids\rightarrow cilia}^{i}=-\frac{(\tau_{m}\boldsymbol{F}_{m}^{i} + \tau_{PCL}\boldsymbol{F}_{PCL}^{i})\boldsymbol{\cdot} \boldsymbol{V}_{i}^{s}}{(\tau_{m}+\tau_{PCL})\| \boldsymbol{V}_{i}^{s} \|^{2}}\boldsymbol{V}_{i}^{s}
\end{align}

\item For each Lagrangian point $s$, the norm of the torque of $\boldsymbol{F}_{fluids \rightarrow cilia}^{i}$ with respect to the base point $O$ of the $i^{\text{th}}$ cilium is computed by: 
\begin{align} 
\| \mathcal{M}_{O}^{i}(\boldsymbol{F}_{fluids \rightarrow cilia}^{i}) \| = \| \boldsymbol{F}_{fluids \rightarrow cilia}^{i} \| L_{p} 
\end{align} 
with $L_{p} = \frac{\|\boldsymbol{X}_{p} \otimes \boldsymbol{V}_{i}^{s} \|}{\| \boldsymbol{V}_{i}^{s} \|}$ being the lever arm. The total fluid feedback onto the $i^{\text{th}}$ cilium considered is then computed by summing the computed quantities over all Lagrangian points:
\begin{align}
T^{i}=\sum_{s=2}^{N_{segments}} \| \boldsymbol{F}_{fluids \rightarrow cilia}^{i} \| L_{p}
\end{align}

The velocity of each cilium is finally modified as follows:
\begin{align}
\|\boldsymbol{V}_{i}\| = \|\boldsymbol{V_{0}}\|+\alpha T^{i}
\end{align}
where $\boldsymbol{V}_{0}$ is the initial speed of the cilia, and $\alpha$ a coupling parameter allowing the tuning of the feedback strength. 

\item Then, at the beginning of the next iteration, the beating pattern of the $i^{\text{th}}$ cilium is adjusted:
\begin{align}
N_{i}^{t} =\mod(N_{i}^{t-1}+\|\boldsymbol{V}_{i}\|,N_{i}^{total})
\end{align}
where $N_{i}^{total}$ is the total number of snapshots defining the beating pattern of the cilia, $N_{i}^{t-1}$ the previous snapshot, and $N_{i}^{t}$ the new position of the $i^{\text{th}}$ cilium (in terms of number of snapshots) at the current iteration.
\end{enumerate}

\subsection{Injection of passive tracers}\label{ssec:tracers}

\begin{figure}
\centering
\includegraphics[scale=0.4]{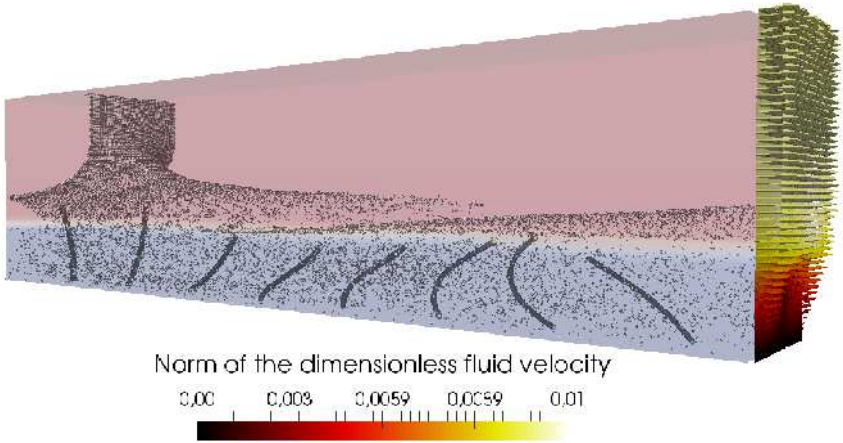}
\caption{Domain filled with passive tracers. The present case corresponds to an antipleptic MCW with $\Delta \Phi = \upi/4$. An array of $8$ cilia with a cilia spacing $a/L=1.67$ is considered on a computational domain of size (N$_{x}=201$, N$_{y}=26$, N$_{z}=50$). The mucus phase (in red) is located above the PCL phase (in blue), and the ratio of viscosity is set such as $r_{\nu}=10$. The colorbar represents the dimensionless velocity magnitude of the fluids on the periodic boundary over the $x$-direction.}
\label{Figure3}
\end{figure}

In the context of real epithelial systems in human body, mucus acts as a barrier against particles and pollutants. To gain insight on how particles are dispersed and advected in the mucus and PCL layers, passive tracers are injected into the domain on a ($y$,$z$) plane, when the flow has reached an established regime (see figure \ref{Figure3} for a view of a domain filled with tracers). Their displacements are then computed and averaged over several cilia beating cycles. Their equations of motion are solved by a second-order Runge-Kutta (RK2) scheme, using the interpolated fluid velocity at each time step and the same procedure as in the IB method.

By taking a closer look at figure \ref{Figure3}, one can observe that particles initially seeded into the PCL are greatly mixed while staying in the PCL. On the contrary, particles initially seeded into the mucus layer stay in the mucus layer and are not mixed. This is mainly due to the surface tension effects present at the mucus-PCL interface which prevent a too strong mixing at the interface. This shows how wetting particles that are deposited to the air-mucus interface might enter the mucus layer but never reach the PCL one.

\section{Results \& Discussion}\label{sec:Result}

\subsection{Emergence of metachronal waves}\label{ssec:MCW}

Using the feedback force introduced in \S \ref{par:feedback}, randomly beating cilia are observed to synchronize with their immediate neighbours giving birth to symplectic or antipleptic metachrony. The time for the synchronization to occur depends on the set of parameters used. Both local and global synchronizations can be observed.

The empirical parameter $\alpha$ plays a role in the emergence of the waves. 
When $\alpha$ is set to a too low value ($|\alpha|<0.5$), the cilia will beat randomly, while if set to a too high value ($|\alpha|>7$ for the case $a/L=1.67$ for example), the cilia will fully synchronize with each other without any phase lag. Additionally, higher absolute values of $\alpha$ usually decrease the times needed for metachronal waves to emerge.

\begin{figure}
\centering
\begin{tabular}{c}
\includegraphics[scale=0.45]{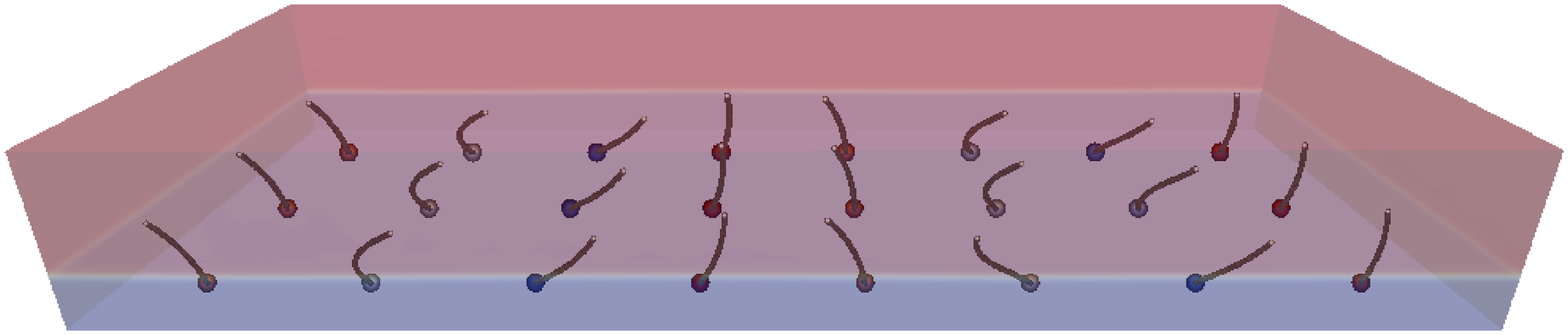}\\
\includegraphics[scale=0.45]{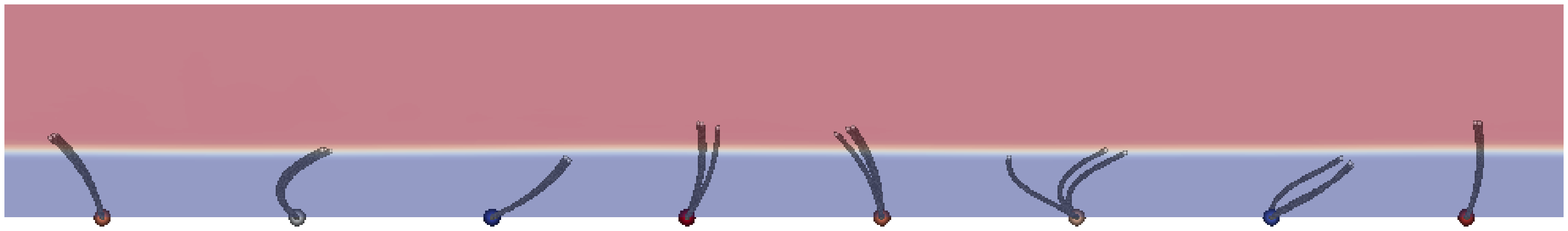}\\
\vspace{0.2cm}\includegraphics[scale=0.2]{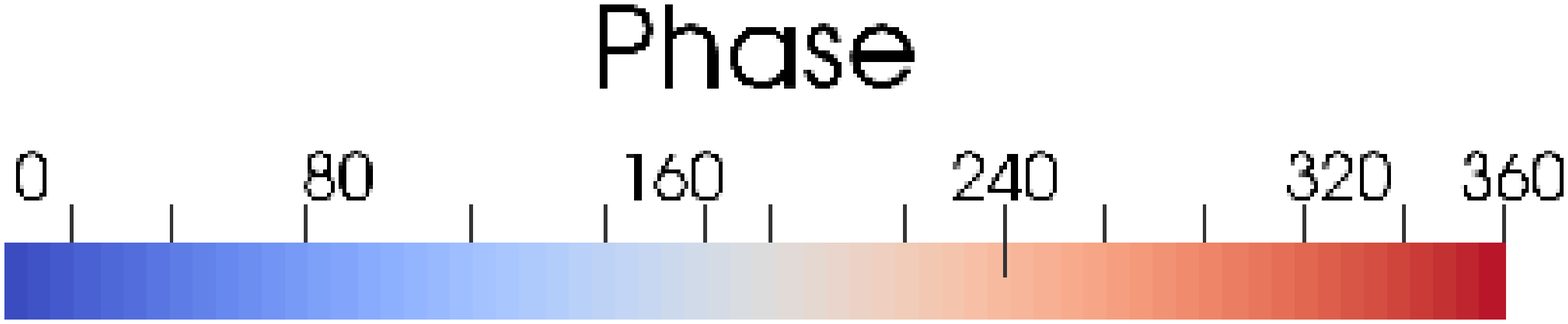}
\end{tabular}
\caption{Symplectic MCW emerging from an initially random state of the cilia. $24$ cilia arranged in a $8\times3$ rectangle are considered on a computational domain of size (N$_{x}=361$, N$_{y}=136$, N$_{z}=50$) with a cilia spacing $a/L=3$. The mucus phase is in red and the PCL phase in blue. The color bar indicates the phase of a particular cilium within one beating period, which is represented by a circle at its base. Top: 3D view of the system. Bottom: 2D view of the same system in a ($x$,$y$) plane to highlight the 3D modulation in the $z$-direction.}
\label{Figure4_5}
\end{figure}

Figure \ref{Figure4_5} shows a symplectic MCW emerging from an initially random state (every cilium were initially beating at a random step of their beat cycle). In the present case, $3$ rows of $8$ cilia on a computational domain of size (N$_{x}=361$, N$_{y}=136$, N$_{z}=50$) are considered. The PCL is set such as $h=0.9L$, and the ratio of viscosity between the PCL and the mucus phase is $15$. The feedback coefficient $\alpha$ used is $\alpha=-3.5$. The spacing between two neighbouring cilia is $a/L=3$ in the $x$ and $y$ directions with $L=15$ \textit{lu}. Hence cilia never collide and overlapping of kernels from neighbouring cilia never occurs. One can clearly see the symplectic MCW that emerged from the initially random state of the cilia, with a wavelength $\lambda=180$ \textit{lu}, and a phase lag $\Delta \Phi \approx -\upi/2$, confirming that hydrodynamic interactions only suffice to account for the emergence of MCW.

\begin{figure}
\centering
\begin{tabular}{c}
\includegraphics[trim={0 100 0 0},clip,scale=0.45]{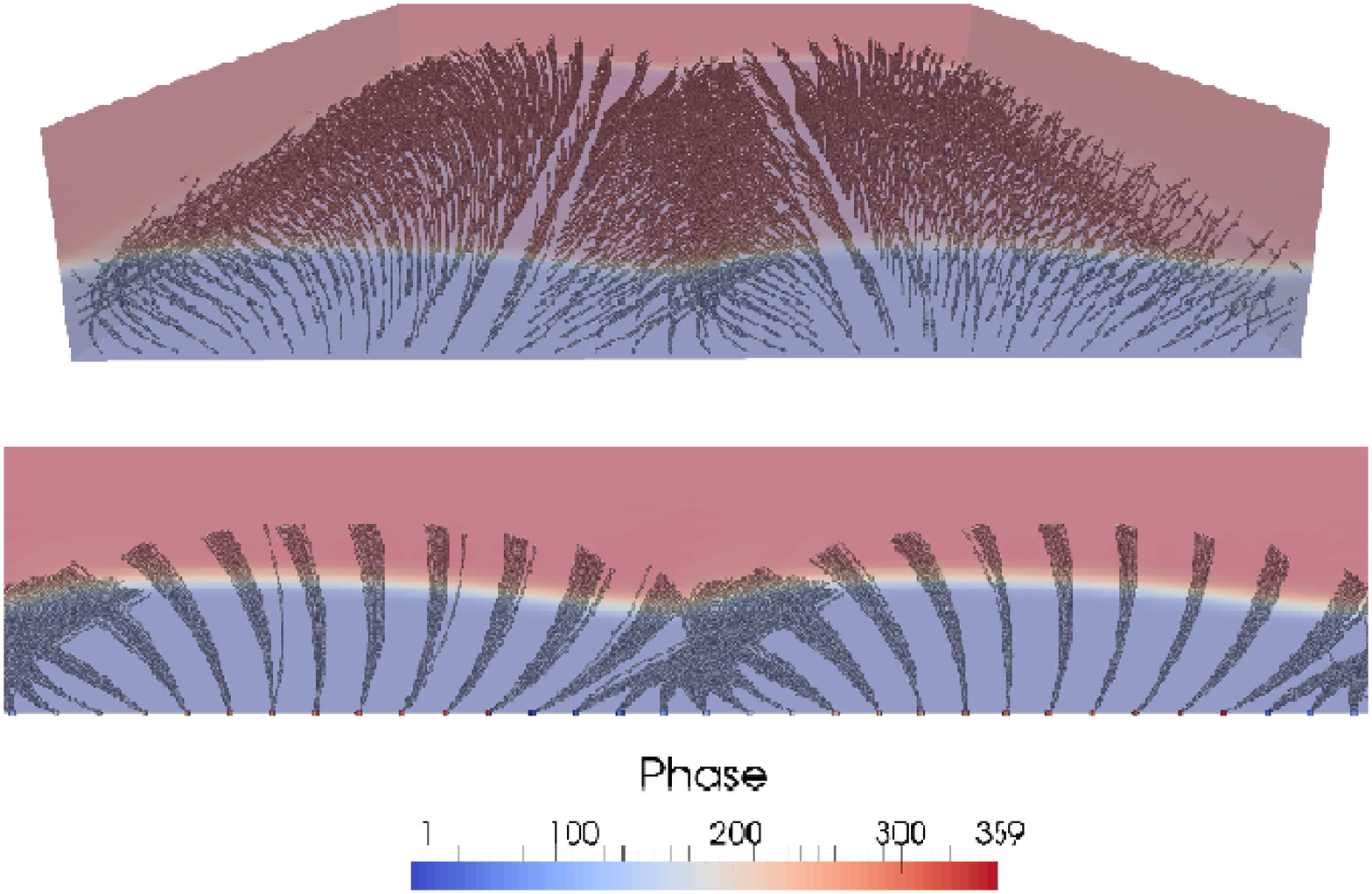}\\
\includegraphics[scale=0.2]{./Figure4_5c.eps}
\end{tabular}
\caption{Antipleptic MCW emerging from an initially random state of the cilia. $1024$ cilia arranged in a $32\times32$ square are considered on a computational domain of size (N$_{x}=161$, N$_{y}=161$, N$_{z}=32$) with a cilia spacing $a/L=0.23$. The mucus phase is in red and the PCL phase in blue. The color bar indicates the phase of a particular cilium within one beating period, which is represented by a circle at its base. Top: 3D view of the system. Bottom: 2D view of the same system in a ($x$,$y$) plane to highlight the 3D modulation in the $z$-direction.}
\label{Figure4}
\end{figure}

Figure \ref{Figure4} shows a similar configuration ($h=0.60L$, $r_{\nu}=15$, $\alpha=-3.5$) where $1024$ cilia arranged in a $32\times32$ square are considered on a computational domain of size (N$_{x}=161$, N$_{y}=161$, N$_{z}=32$). 
One can clearly see the formation of an antipleptic MCW with a wavelength $\lambda=80$ \textit{lu}, and a phase lag $\Delta \Phi\approx \upi/8$. Notice that in the simulation presented in figure \ref{Figure4} the cilia spacing has been reduced ($a/L=0.23$ and $L=22$ \textit{lu}), to take into account a larger number of cilia. Thus, neighbouring cilia may overlap, which can cause spurious numerical effects, due to the 2D beating pattern of the cilia, when two Lagrangian points with opposite velocities occupy the same mesh point. 
However, it has been verified on some simulations, that these effects do not play any role in the emergence of the MCW by setting to zero the IB forces contribution of these Lagrangian points, therefore canceling the spurious effects that may occur at the overlapping points of neighbouring cilia. In other words, the fluid velocity at the Eulerian node surrounding these particular Lagrangian points was not modified by the IB method, proving that it is not the source of the cilia synchronization.
As shown in figures \ref{Figure4_5} and \ref{Figure4}, weakly 3D effects may be observed but do not play a key role in the physics of emerging waves. In all the simulations presented, the synchronization along the $x$-direction is quite strong. In some simulations, particular cilia may lose synchronization over time, but quickly readjust their beating accordingly to the others. It also appears that once an equilibrium is reached, the synchronization along the $y$-direction is stable too. A remarkable fact is that for particular configurations (as the one displayed in figure \ref{Figure4}), the interface between the mucus and the PCL also forms a wave traveling in the same direction as the MCW. If the mucus viscosity is increased, the torques felt by the cilia become stronger. Hence, the cilia are either more accelerated ($\alpha>0$) or decelerated ($\alpha<0$), and the mucus will adapt its displacement accordingly. 
However, when the stationary regime is reached, the clearance velocity should not be strongly modified as shown by \citet{ChatelinPOncet}. According to these authors, the mucus velocity is decreased by only 50\% when the viscosity ratio is increased by a factor of 100000.
Preliminary results (not shown) seem to indicate that, assuming a least effort behaviour of the cilia (meaning $\alpha<0$), antipleptic waves are obtained for small cilia spacings ($a/L \le 1.5$) while symplectic waves are seen to emerge for higher cilia spacings ($a/L \ge 1.5$). Since in nature, antipleptic waves are often observed for densely packed cilia, it implies that natural cilia adopt a least effort behaviour. It also suggests the existence of a critical value for the cilia density, $\rho_{c}=\rho(\alpha_{c})$ where $\alpha_{c}$ is the corresponding critical value of the coupling parameter, from which the kind of waves emerging can be controlled. On the contrary, with the present model, assuming that the cilia beat faster when encountering a resistance (meaning $\alpha>0$), symplectic waves are seen to emerge for small cilia spacings ($a/L \le 1.5$) while antipleptic waves emerge for higher cilia spacings ($a/L \ge 1.5$), which is not observed in nature. Movies of the MCW are provided as online supplementary materials.

%

As already explained in \S \ref{ssec:geometrical}, when the feedback is taken into account, a non symmetrical motion develops, with a stroke phase slower than the recovery phase for both antipleptic and symplectic MCW. For the antipleptic case, $\|\boldsymbol{V}_{i}\|$ is smaller during the stroke phase, and since $\alpha$ is negative and the torques $T^{i}$ are always positive, it means that the values of $T^{i}$ computed during the stroke phase are larger than the values computed during the recovery phase. It results in a weaker velocity for the cilia which cover less snapshots during the stroke phase.
It is the opposite for the symplectic case where $\alpha$ is positive: $T^{i}$ takes larger values during the recovery phase as $\|\boldsymbol{V}_{i}\|$. 
In other terms, the feedback depends on the clustering character of the motion: when cilia are clustered, the torque exerted by the fluids onto the cilia is weaker, and when they are far from each other (during the stroke phase for antipleptic motion and during the recovery phase for symplectic motion), the torques are stronger. This makes sense since the cilia encounter more viscous resistance when they are not clustered, as it will be discussed further down in \S \ref{ssec:NumPower}. The resulting motion of the cilia then differs from what is observed in nature, but it indicates that cilia experience stronger stresses during the stroke phase of antipleptic motion and during the recovery phase of symplectic motion. It supposes that the beating kinematics of the real cilia are not dictated only by hydrodynamical interactions and suggests that other biological parameters or functions (such as sensing) may play a role. This is in agreement with \citet{GuoH} who compared the performance of pumping-specialized cilia and swimming-specialized cilia as a function of the metachronal coordination, and found that the later almost always outperforms the pumping-specialized cilia. As it will be further detailed in \S \ref{ssec:displacement_ration}, their results are also in accordance with the outcome of the present paper: antipleptic waves are the most efficient ones to transport fluids. Finally, it is worth noticing that the degree of asymmetry is much higher for antipleptic MCW compared to symplectic MCW.

\subsection{Quantitative study of the metachronal waves}\label{sec:Study}
\begin{figure}
	\centering
	\begin{tabular}{c}
	\includegraphics[scale=0.45]{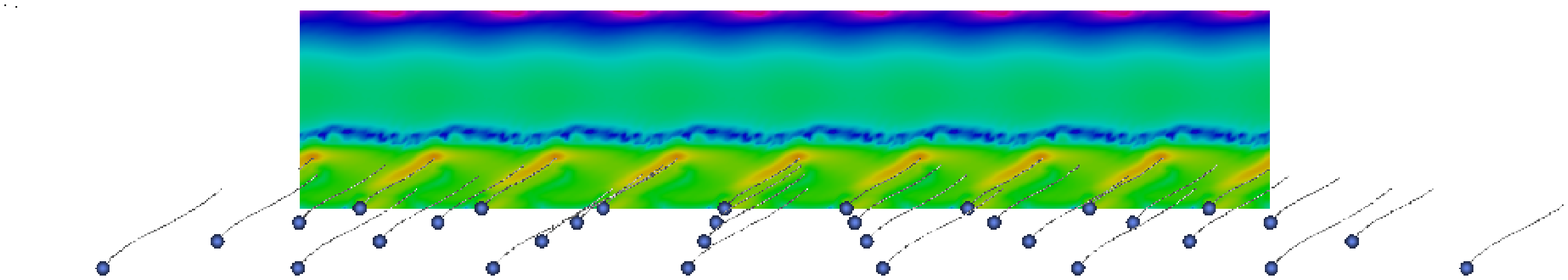}\\
	(a) $\Delta \Phi = 0$\\
	\includegraphics[scale=0.45]{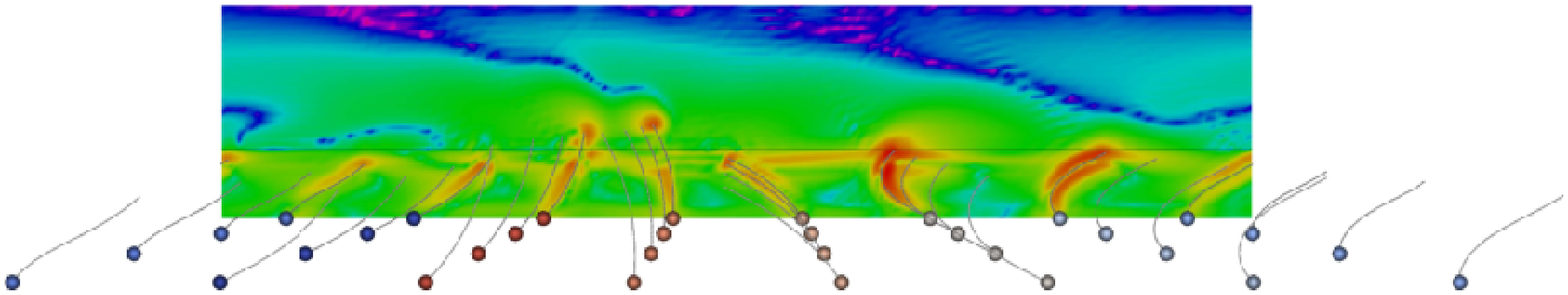} \\
	(b) $\Delta \Phi = -\upi/4$\\
    \includegraphics[scale=0.45]{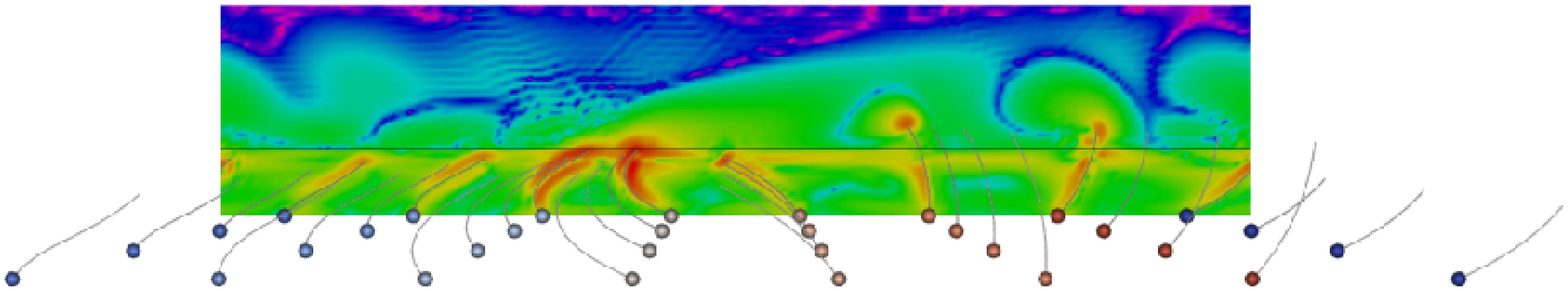} \\
	(c) $\Delta \Phi = \upi/4$\\
	\includegraphics[scale=0.45]{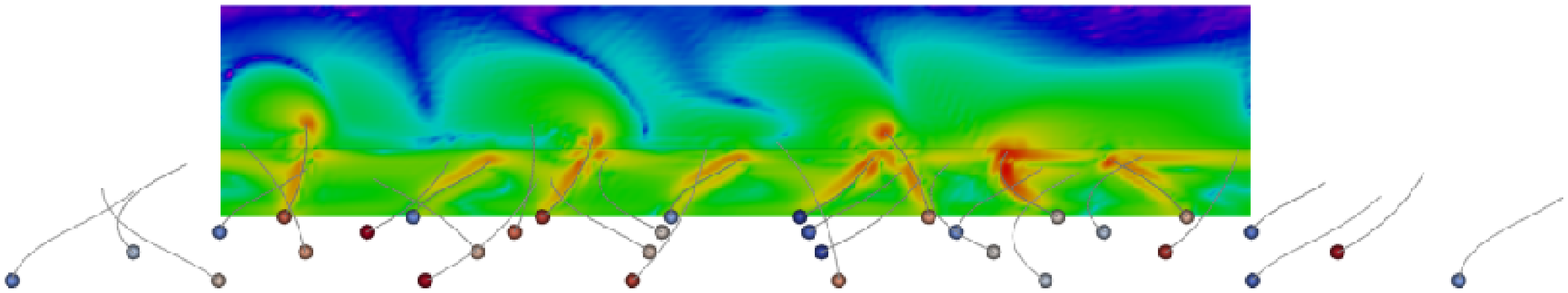}\\
	(d) Random \\
	\includegraphics[scale=0.3]{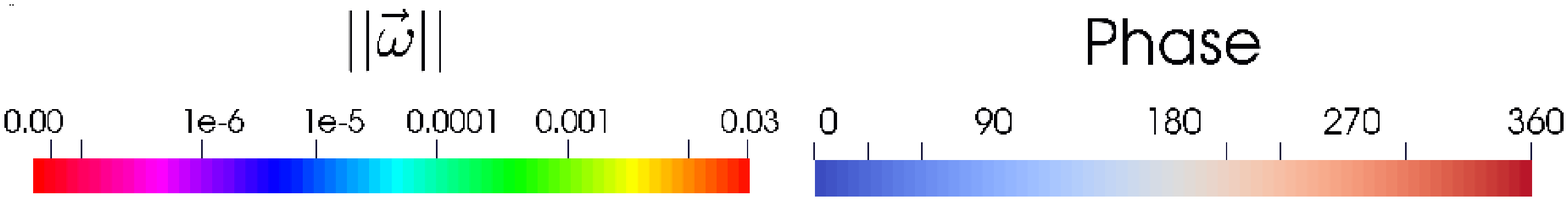} \\
	\end{tabular}
	\caption{Different kinds of collective coordination for the beating cilia. $32$ cilia arranged in a $8\times4$ square are considered on a computational domain of size (N$_{x}=241$, N$_{y}=121$, N$_{z}=50$) in each case, with a cilia spacing $a/L$=2. (a) Synchronized motion; (b) Symplectic metachronal motion; (c) Antipleptic metachronal motion; (d) No synchronization (random state of cilia). The figures show contours of the magnitude of the dimensionless vorticity $\| \vec{\omega} \|$ using a logarithmic scale. The black lines show the frontier between the PCL at the bottom and the mucus layer above.}
	\label{Figure5}
\end{figure}

This section presents the results obtained, once the flow is well-established, for the three configurations studied: (a) synchronized case (all cilia beat together with no phase lag); (b) symplectic MCW where two neighbouring cilia beat with a negative phase lag, i.e $-\upi <\Delta \Phi <0$; and (c) antipleptic MCW where two neighbouring cilia beat with a positive phase lag, i.e $0< \Delta \Phi < \upi$. Note that in all following results, contrary to \S \ref{ssec:MCW}, the metachrony is imposed in order to study specific phase lags $\Delta \Phi$. Hence, the size of the domains must be changed accordingly, since different phase lags $\Delta \Phi$ imply different wavelengths, and hence more or less cilia. However, the quantities presented here have been averaged over the domain, and there are no effects in changing the size of the box. In order to study only spatially asymmetric motions, the recovery and stroke phases now take the same amount of time in all following results. Also note that the standard deviations are not displayed in any of the figures that will be presented as they are extremely small (less than 0.01$\%$) for all considered quantities. Hence they do not give additional information. Schematic views of synchronized beating and metachronal motions are displayed in figure \ref{Figure5}. Another configuration, corresponding to randomly beating cilia is also represented in figure \ref{Figure5}. 
One can observe that the synchronized motion of cilia creates vorticity only in the periciliary zone, 
whereas metachronal motions also induce vorticity into the mucus layer. 
It is worth noticing that for symplectic motion (case (b)), vortex trails emerge from the cilia tips performing their stroke phase. The same phenomenon is observed in the antipleptic case during the recovery motion of cilia. The presence of coherent vortices is clearly visible in both cases.

\subsubsection{Transport and mixing zone}\label{sssec:transport_and_mix}

\begin{figure}
\centering
\begin{tabular}{cc}
\includegraphics[scale=0.42]{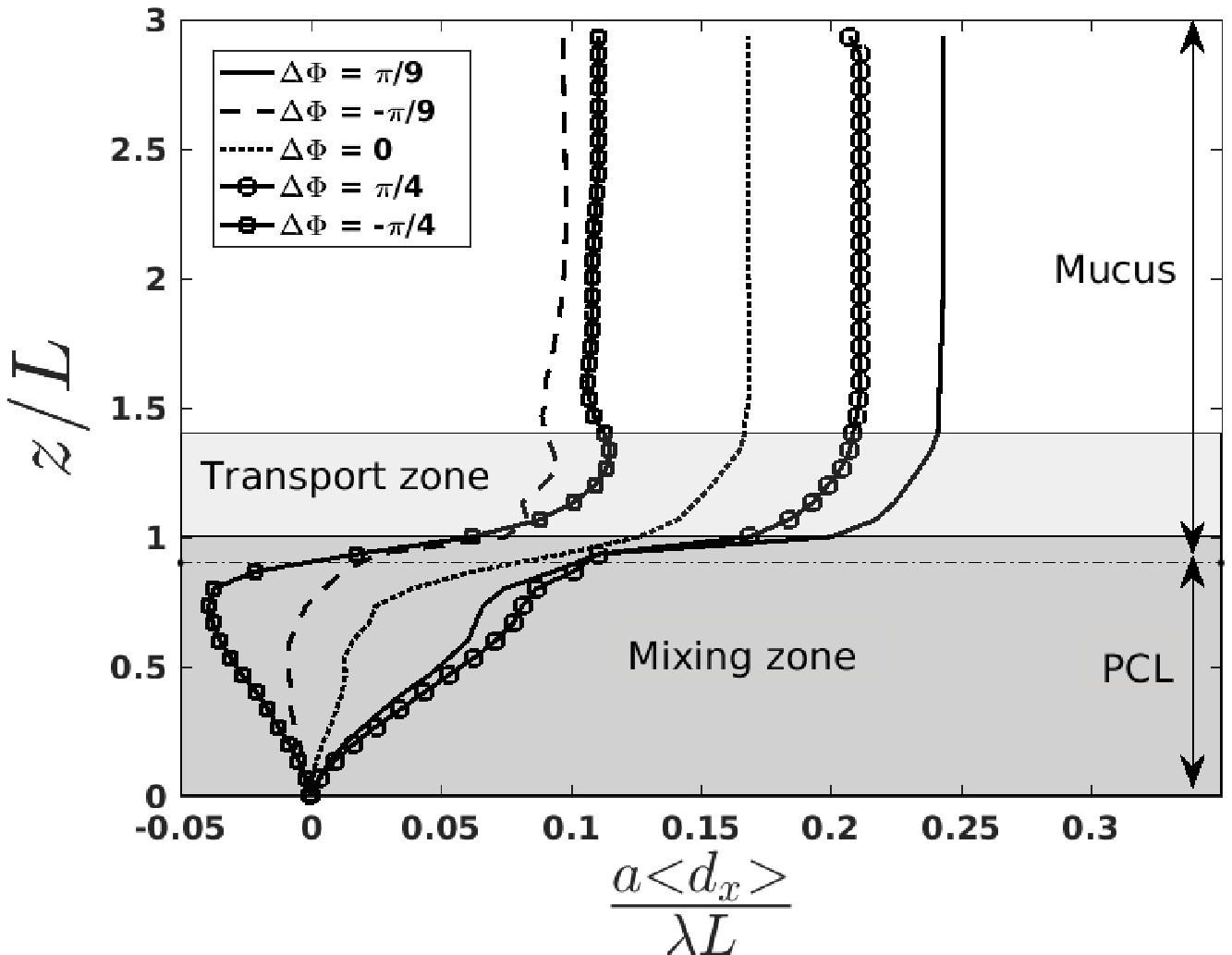} &
\includegraphics[scale=0.42]{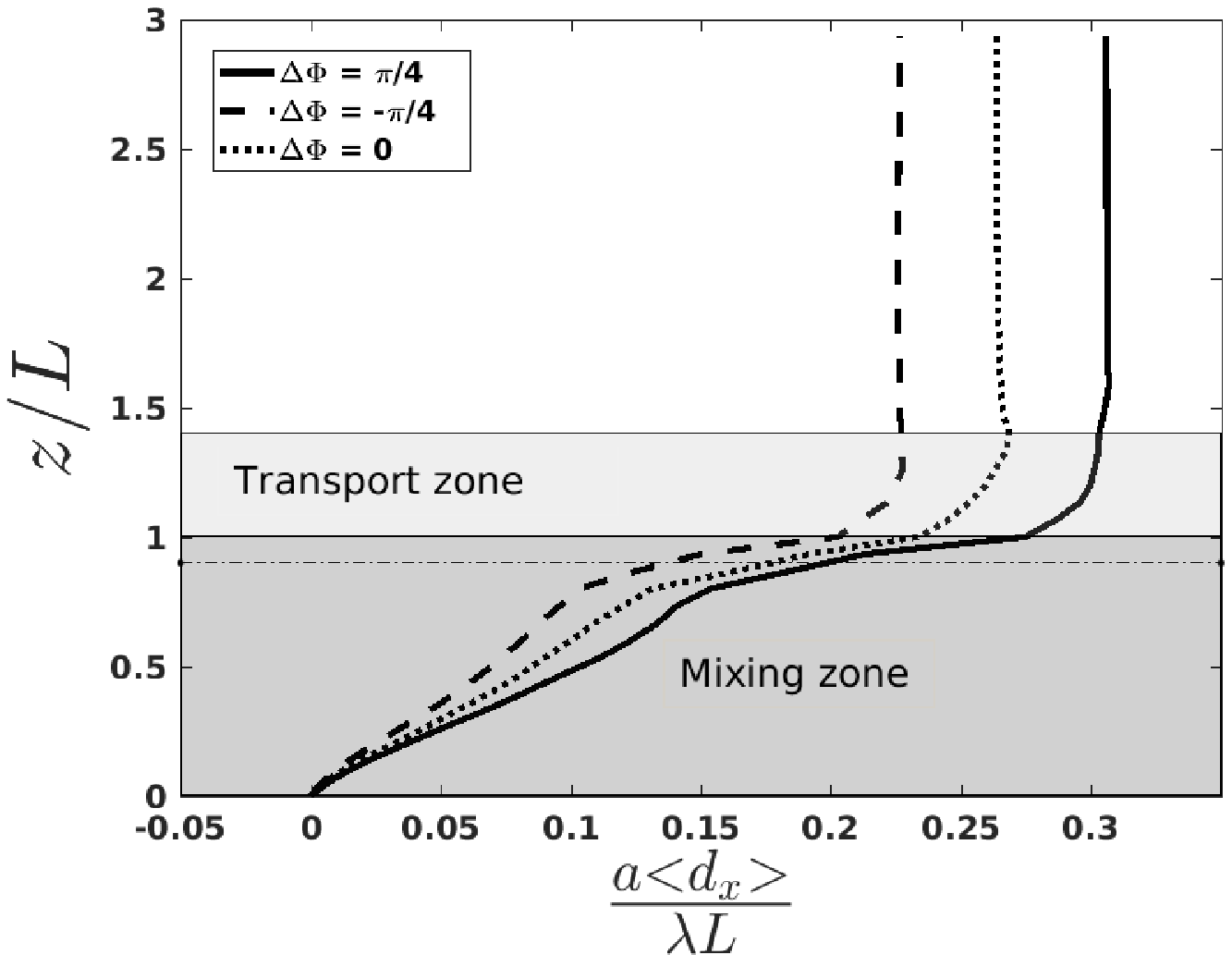} \\
(a) & (b)
\end{tabular}
\caption{Normalized average displacement in the $x$-direction as a function of $z/L$ for (a) $\Delta \Phi = \pm \upi/9$ and $\Delta \Phi = \pm \upi/4$ with $a/L=0.8$; (b) $\Delta \Phi = \pm \upi/4$ and $a/L=1.67$. As in \citet{Ding}, the mixing zone extends from 0 to $1L$, and the transport zone from $1L$ to $1.4L$. In both cases (a) and (b), the mucus-PCL interface is located at $z/L=0.9$ and is indicated by a dashed horizontal line.}
\label{Figure6}
\end{figure} 

The displacement field is calculated such as $\boldsymbol{d}(\boldsymbol{x})$$=$$\int_{0}^{T}\boldsymbol{u}(\boldsymbol{x}(t),t)$dt, where $\boldsymbol{x}$ is the position vector, and $\boldsymbol{u}$ the fluid velocity.
Its component over the $x$-direction as a function of the axial position is averaged over several beating cycles and displayed in figure \ref{Figure6}. The \textit{transport} and \textit{mixing} zones defined in \citet{Ding} are clearly visible. 
It shows that the antipleptic MCW perform better than the other kind of coordination to transport the particles.
Note that in figure \ref{Figure6}b, the two curves with $\Delta \Phi = \pm\upi/4$ have a similar shape, whereas in figure \ref{Figure6}a symplectic MCW induce a negative transport (i.e. a counter-flow) in the \textit{mixing zone} for the same phase lag. For a given value of phase lag, the transport depends on the density of cilia. 
One can also notice the importance of the phase lag by looking at figure \ref{Figure6}a: for a phase lag $\Delta \Phi=-\upi/9$, the symplectic MCW induce no counter-flow in the mixing zone while the opposite happens for $\Delta \Phi=-\upi/4$.
Finally, it is worth noticing that, as the cilia spacing increases, the influence of the kind of metachrony decreases. Eventually, for very large cilia spacings, one expects that the 3 curves displayed in figure \ref{Figure6}b would merge into a single one corresponding to the displacement generated by an isolated cilium. Additionally, even with a larger cilia spacing, the particles are better transported in the case (b).
As in \citet{Ding}, the displacement reaches a plateau for an axial position of $1.4L$ in both cases (a) and (b). The beating pattern used in \citet{Ding} is different, but similar trends are observed. Nevertheless, a major difference must be pointed out: for $\Delta \Phi = -\upi /4$ and $a/L=1.67$ (figure \ref{Figure6}b), the present results show that the displacement induced by the symplectic wave has the same shape as the synchronized and antipleptic cases, i.e it induces transport in the mucus phase even if it is smaller. 
On the contrary, \citet{Ding} obtained no transport at all for the symplectic wave in the mucus phase, but instead a peak of transport under the cilia tips in the PCL at $z=0.7L$ due to the presence of a vortex-like structure both below and above the cilia tips for this particular value of the phase lag $\Delta \Phi$. Since the main difference between both studies is the beating pattern of the cilia (same phase lag and cilia spacing), it highlights the sensitivity of the system to the beating pattern.

The displacement in the $z$ and $y$-directions have also been analyzed. The results show that the $y$-component of the displacement could be neglected and that the $z$-component of the displacement was almost null above the cilia tips, but not zero under. As it will be discussed later, the $z$-component of the displacement reaches its maximal value in regions where the shear rate is maximal too. It agrees with the existence of a mixing zone under the cilia tips.

\begin{figure}
\centering
\begin{tabular}{c}
\includegraphics[scale=0.4]{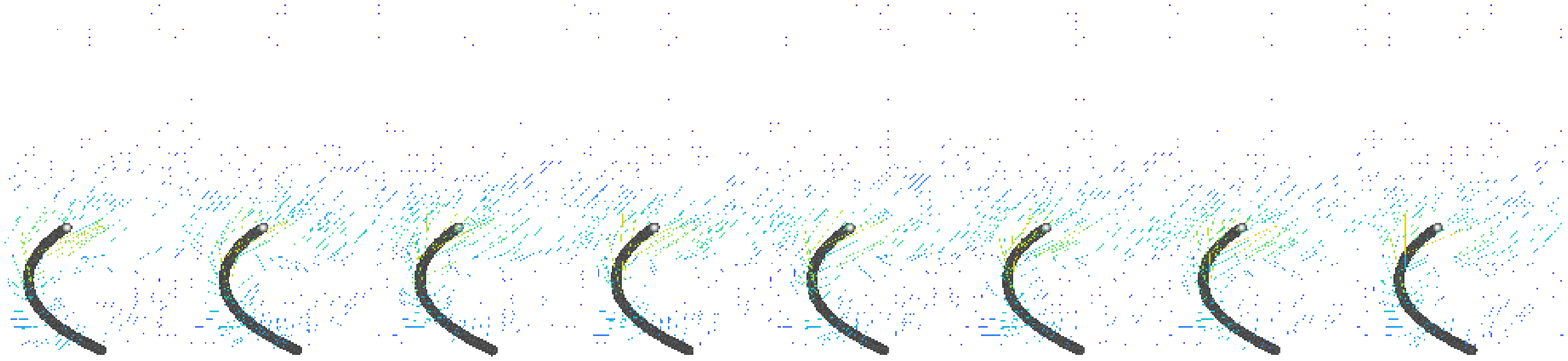} \\
(a) \\
\includegraphics[scale=0.4]{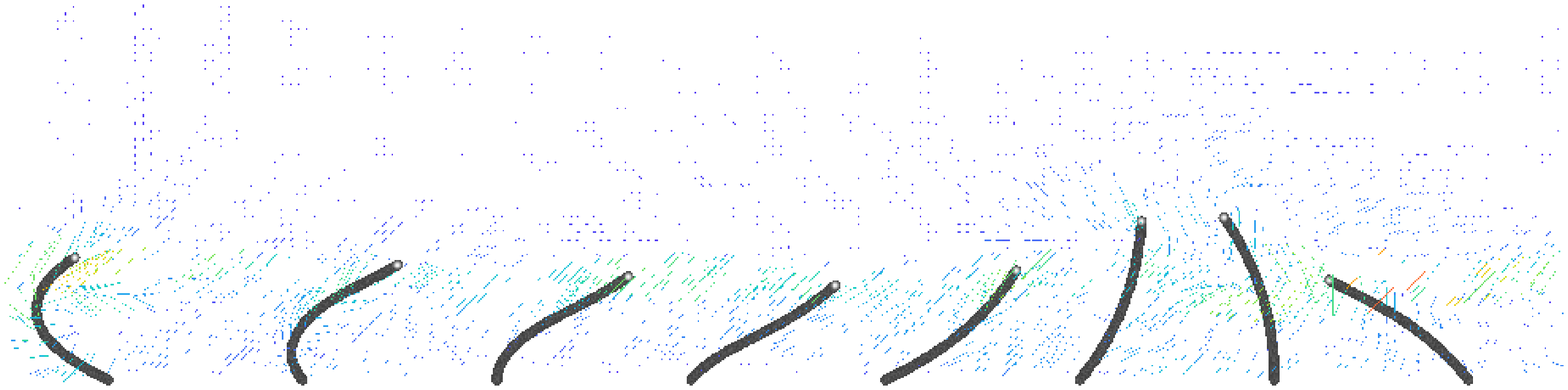} \\
(b) \\
\includegraphics[scale=0.4]{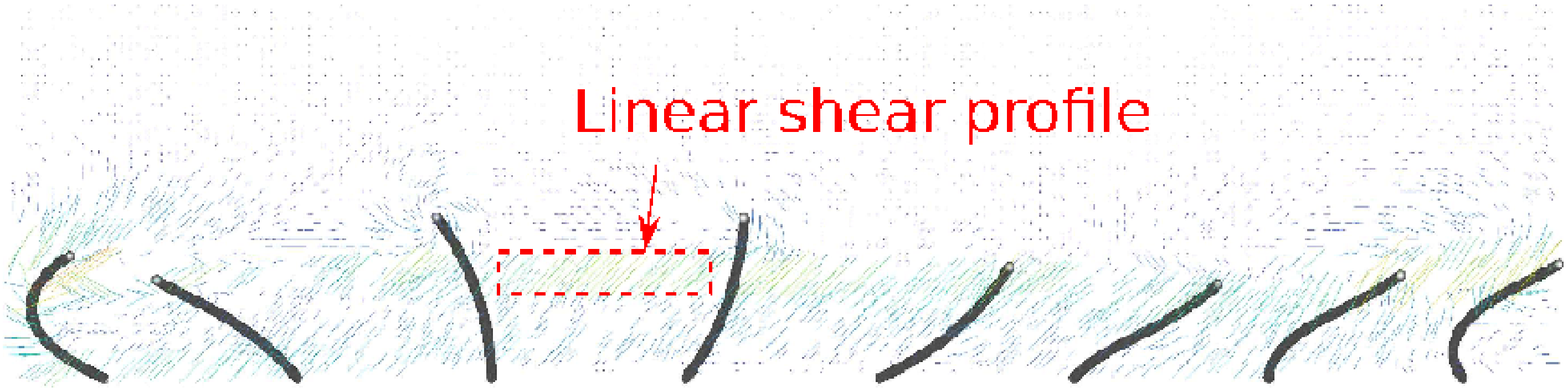} \\
(c) \\
\includegraphics[scale=0.04]{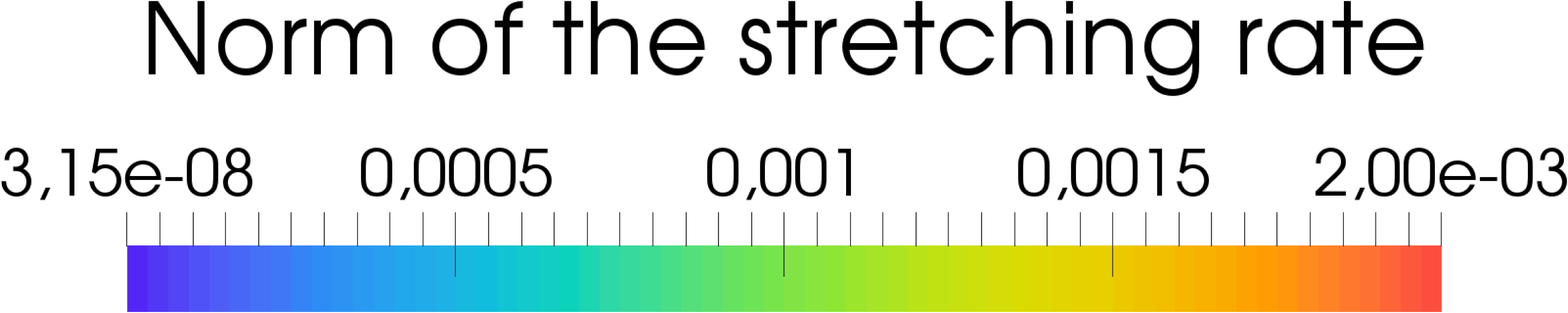} \\
\end{tabular}
\caption{Stretching rate and direction for: (a) Synchronized motion ($\Delta \Phi=0$), (b) Symplectic MCW ($\Delta \Phi=-\upi/4$), (c) Antipleptic MCW ($\Delta \Phi=\upi/4$). In each case, the cilia spacing is $a/L=1.67$, and the size of the computational domain is (N$_{x}=201$,N$_{y}=26$,N$_{z}=50$).}
\label{Figure7}
\end{figure}

By doing an analogy with the strain-rate tensor classically used in solid mechanics, the gradient of the displacement field $\nabla d$ can be computed by considering each Eulerian node of the Cartesian grid as passive tracers. Since the displacement in the transverse direction is very small, the $y$-component of the displacement field can be neglected. Henceforth, the gradient is computed over every ($x$,$z$) plane, and an average is then done over the $y$-direction. Following the methodology described in \citet{Ding}, $\nabla d$ is decomposed into an antisymmetric component $\mathsfbi{R}=(\nabla d - (\nabla d)^{T})/2$ corresponding to rotation, and a symmetric component $\mathsfbi{S}=(\nabla d + (\nabla d)^{T})/2$ corresponding to shear deformation. The two eigenvalues of $\mathsfbi{S}$ are of the form $\pm \gamma$ and indicate the rates of stretching ($+ \gamma$) and compression ($- \gamma$). The unit eigenvector $\boldsymbol{e}_{\gamma}$ corresponding to the positive eigenvalue indicates the direction of stretching, and the other eigenvalue the direction of compression. Because of the incompressibility condition, both eigenvectors are orthogonal, and so plotting only one of them is sufficient to have the complete set of information. In figure \ref{Figure7}, the stretching rate and its direction are presented in a ($x$,$z$) plane for the three cases studied. The stretching rate is maximal near the upper part, and at little distance above the cilia tips in all three cases. For the synchronized motion (case (a)) there is almost no stretching away from the cilia during the recovery phase whereas for cases (b) and (c), corresponding respectively to symplectic and antipleptic motion, a weak stretching can always be observed (see the right side of cases (b) and (c) in figure \ref{Figure7}).
A complex shape of the stretching rate is observed at the cilia tips during the stroke phase in antipleptic motion (see cilium 3 for instance); during the stroke phase in symplectic motion (see cilia 6 and 7) and during the stroke phase in synchronized motion (results not shown).
From their orientations, one can expect an enhancement in the mixing in this region.
As in \citet{Ding}, the stretching direction is a nonlinear function of space. Except for zones where the shear rate is maximal, the $\gamma \vec{\boldsymbol{e}}_{\gamma}$ field is then oriented at 45$^{\circ}$ compared to the $x$-direction, and almost uniform. This is reminiscent of a linear shear profile $d_{x}=cz$ where $c$ is a constant. The nonuniform aspect of the stretching orientation indicates the presence of ``folding'' in the displacement field $\boldsymbol{d}$ which also plays a role in the mixing as explained by \citet{Kelley}. \citet{Ding} obtained that the stretching rate was maximal for the antipleptic case for $\Delta \Phi=\upi/4$ which is exactly what is obtained here, even if the beating patterns used are different. One can then expect that this value of phase lag $\Delta \Phi$ is the most efficient in mixing fluids by enhancing the stretching near the upper part of the cilia.

In order to assess the reliability and robustness of the solver, a comparison with experimental data is also performed. The average clearance velocity was computed on the plane ($x$,$y$,$3.2L$) for all simulations. The highest clearance velocity is reached for the case $a/L=1.67$ with $\Delta \Phi=\pi/4$ and equals $33.47$ $\mu$m/s. It is in good agreement with the experimental results of \citet{Matsui} who observed a clearance velocity of $39.8\pm4.2$ $\mu$m/s for the mucus. \citet{Matsui} also observed that the PCL flows in the same direction as the mucus, which is the case in the present study. Finally, the PCL-mucus interface remains approximately flat in most of the simulations presented, as it is the case in the micrographs performed by \citet{Sanderson} on rabbit tracheal epithelium.

\subsubsection{Directional pushing efficiency}\label{sssec:Direction}

Inspired by the works of \citet{Khaderi11}, \citet{Gauger2009}, and \citet{KimNetz}, a positive flux $Q_{p}$ and a negative flux $Q_{n}$ are defined as follows: the $x$-component of the velocity field is considered over the (N$_{x}$,$y$,$z$) plane and, at each time step of a full beating cycle, the negative and positive velocity values are separated. The instantaneous negative and positive fluxes are then computed over a sufficiently large number of cycles. The difference $(Q_{p}-Q_{n})$ gives the net instantaneous flux; and the directional pushing efficiency $\epsilon_{PN}$ is defined as : $\epsilon_{PN}=(Q_{p}-Q_{n})/(Q_{p}+Q_{n})$.
Note that here, contrary to the previous works of the aforementioned authors, the domain is not restricted to the \textit{transport area} but covers the whole flow region instead. Such a choice is justified by the fact that it has been experimentally observed, using confocal microscopy and fluorescent markers, that PCL and mucus are in reality transported at approximately the same rate \citep{Matsui}. As a matter of fact, the PCL transport seems to depend on the presence of a mucus layer above it, and ciliary mixing is thought to be responsible for the diffusion of momentum from mucus to PCL. Taking into account the \textit{transport zone} and the \textit{mixing zone} in the computation of $\epsilon_{PN}$ then allows one to collect information about how transport and mixing work together during a beating cycle.

\begin{figure}
\centering
\includegraphics[scale=0.3]{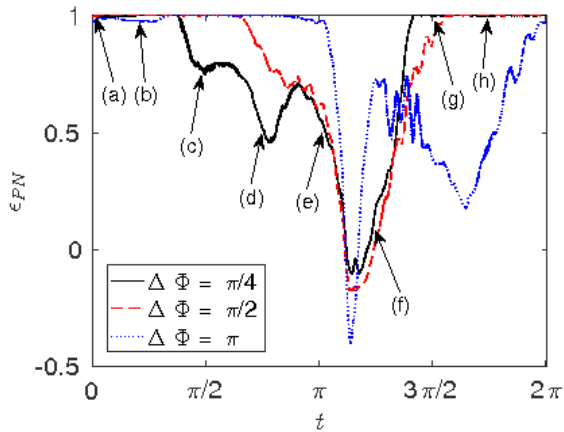}
\caption{Directional pushing efficiency $\epsilon_{PN}$ over one beating cycle for three different phase lags $\Delta \Phi$. Results obtained for an antipleptic MCW and $a/L=2$.}
\label{Figure8}
\end{figure}

The variation of $\epsilon_{PN}$ over one beating cycle for five different phase lags $\Delta \Phi$, all for antipleptic MCW with a cilia spacing such as $a/L=2$, is illustrated in figure \ref{Figure8}. Let us remind that, although the quantities considered here are the results of the motion of all cilia, the calculations have been made through the beating cycle of a single cilium of reference. Therefore, the fact that the corresponding curves have a huge decrease around $t = \upi$ is related to the choice of this particular cilium. Indeed, the plane used to compute the positive and negative fluxes is located close to a particular cilium. This cilium is in the recovery phase at $t=\upi$, inducing a small value of $\epsilon_{PN}$ at this particular time. Choosing another cilium or another plane would shift horizontally the drop observed around $t = \upi$. 

\begin{figure}
\centering
\begin{tabular}{ccccccccc}

(a) $t=0$ & \multicolumn{8}{c}{\hspace{0.34cm}\vspace{-0.2cm}\includegraphics[scale=0.3]{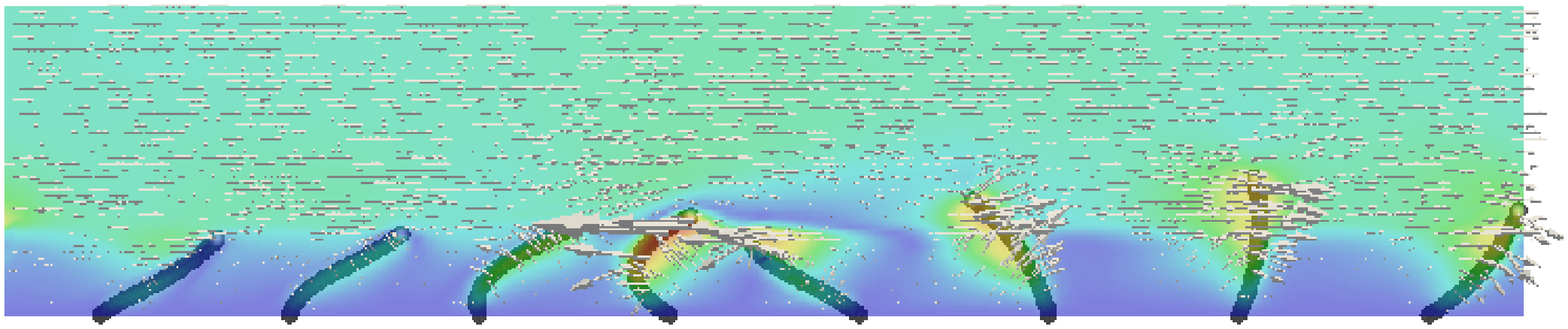}}\\
&\tiny\hspace{0.75cm} 1&\tiny\hspace{0.72cm} 2&\tiny\hspace{0.72cm} 3&\tiny\hspace{0.72cm} 4&\tiny\hspace{0.72cm} 5&\tiny\hspace{0.72cm} 6&\tiny\hspace{0.72cm} 7&\tiny \hspace{0.0cm}8\\

(b) $t=\frac{\upi}{4}$ &\multicolumn{8}{c}{\hspace{0.20cm} \vspace{-0.2cm} \includegraphics[scale=0.3]{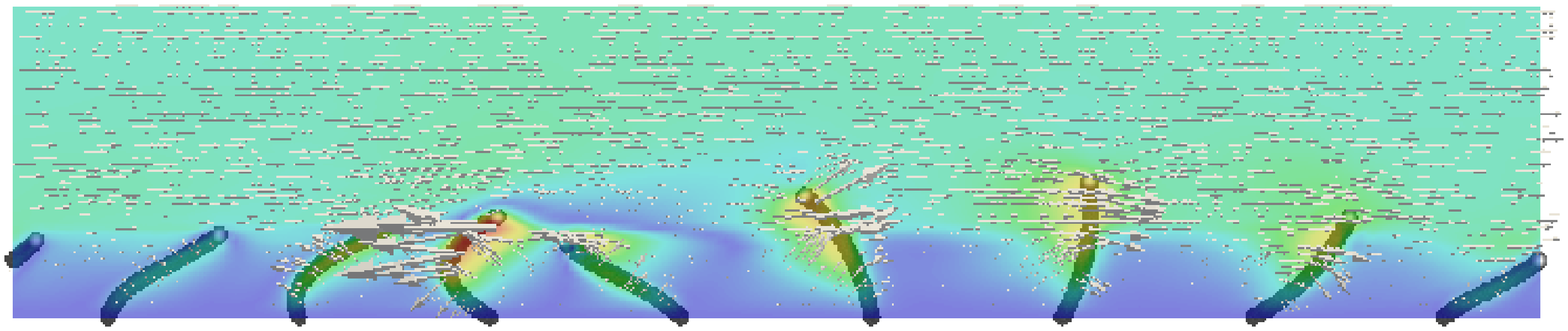}}\\
&\tiny\hspace{0.75cm} 1&\tiny\hspace{0.72cm} 2&\tiny\hspace{0.72cm} 3&\tiny\hspace{0.72cm} 4&\tiny\hspace{0.72cm} 5&\tiny\hspace{0.72cm} 6&\tiny\hspace{0.72cm} 7&\tiny \hspace{0.0cm}8\\
(c) $t=\frac{\upi}{2}$ &\multicolumn{8}{c}{\hspace{0.24cm}\vspace{-0.2cm}\includegraphics[scale=0.3]{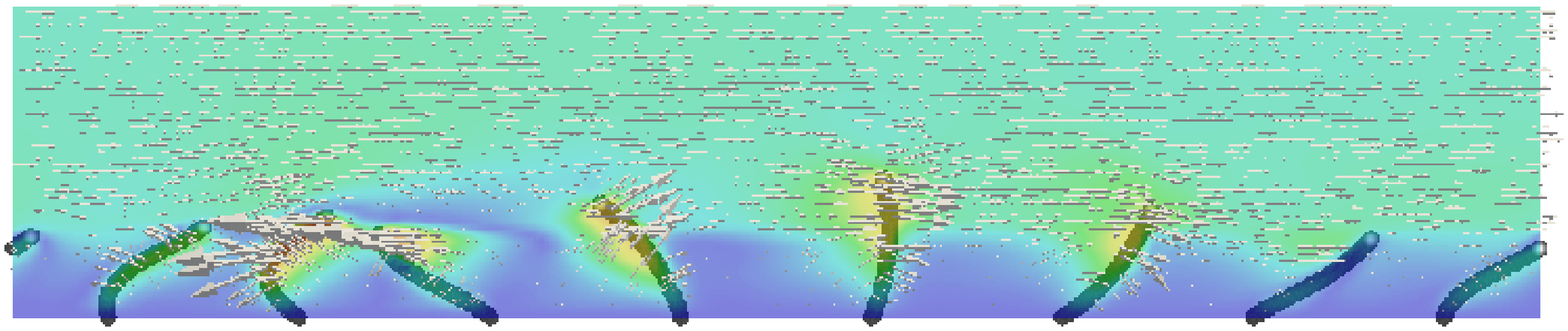}}\\
&\tiny\hspace{0.75cm} 1&\tiny\hspace{0.72cm} 2&\tiny\hspace{0.72cm} 3&\tiny\hspace{0.72cm} 4&\tiny\hspace{0.72cm} 5&\tiny\hspace{0.72cm} 6&\tiny\hspace{0.72cm} 7&\tiny \hspace{0.0cm}8\\
(d) $t=\frac{3\upi}{4}$ &\multicolumn{8}{c}{\vspace{-0.2cm}\includegraphics[scale=0.31]{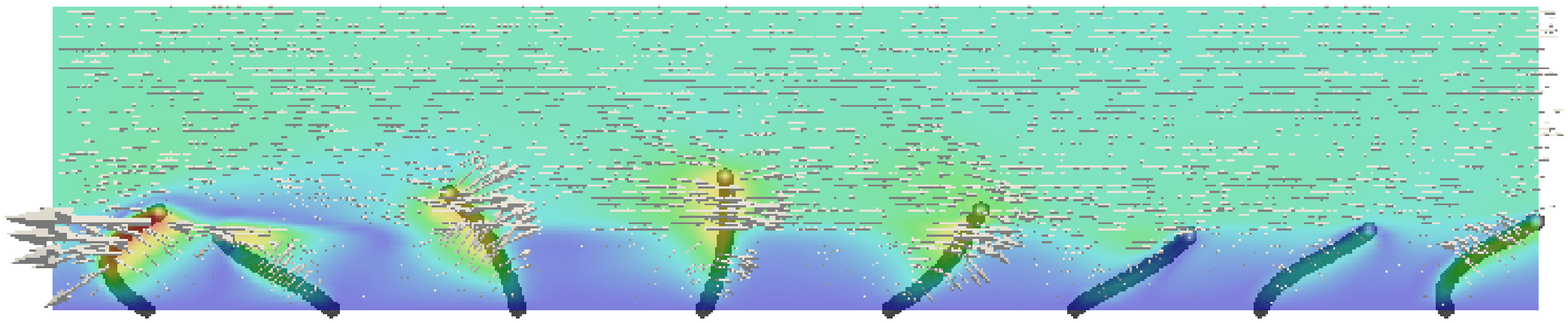}}\\
&\tiny\hspace{0.75cm} 1&\tiny\hspace{0.72cm} 2&\tiny\hspace{0.72cm} 3&\tiny\hspace{0.72cm} 4&\tiny\hspace{0.72cm} 5&\tiny\hspace{0.72cm} 6&\tiny\hspace{0.72cm} 7&\tiny \hspace{0.0cm}8\\
(e) $t=\upi$&\multicolumn{8}{c}{\vspace{-0.2cm}\includegraphics[scale=0.31]{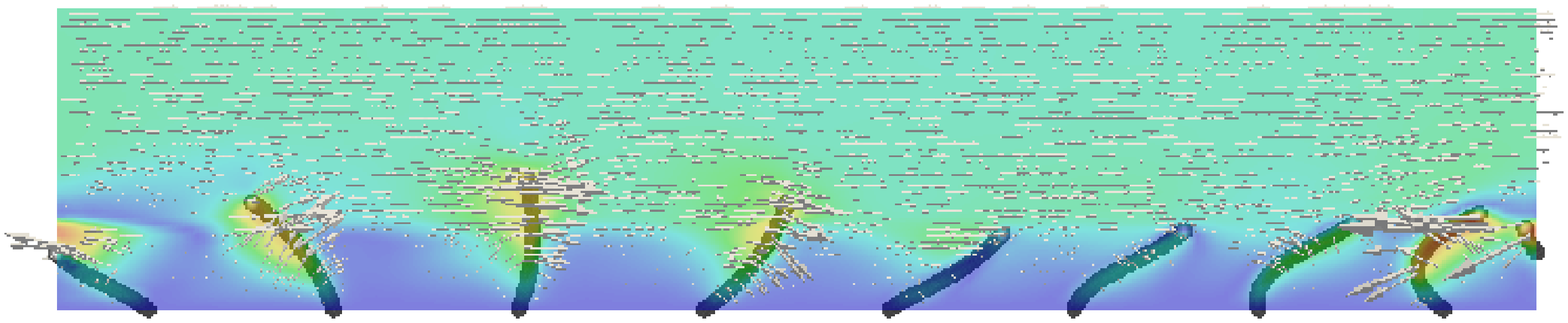}}\\
&\tiny\hspace{0.75cm} 1&\tiny\hspace{0.72cm} 2&\tiny\hspace{0.72cm} 3&\tiny\hspace{0.72cm} 4&\tiny\hspace{0.72cm} 5&\tiny\hspace{0.72cm} 6&\tiny\hspace{0.72cm} 7&\tiny \hspace{0.0cm}8\\
(f) $t=\frac{5\upi}{4}$&\multicolumn{8}{c}{\hspace{0.295cm}\vspace{-0.2cm}\includegraphics[scale=0.3]{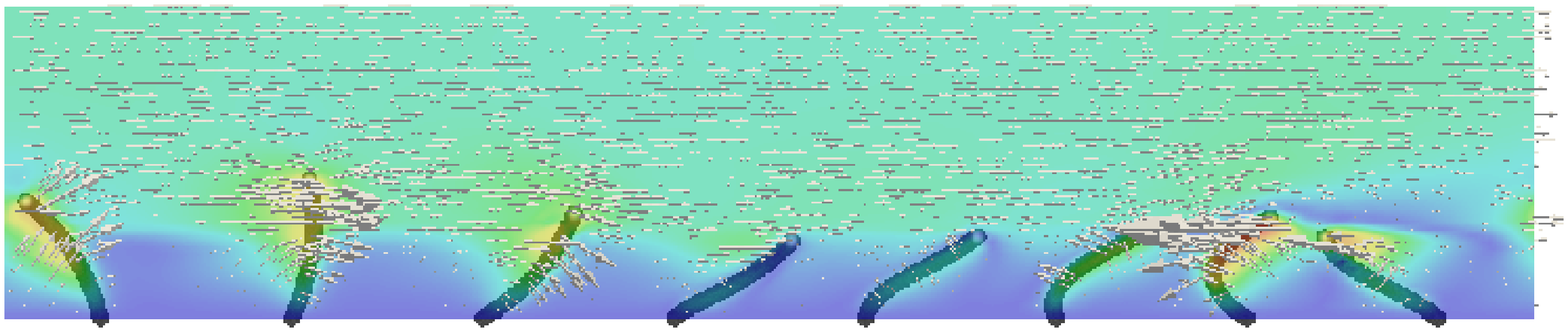}}\\
&\tiny\hspace{0.75cm} 1&\tiny\hspace{0.72cm} 2&\tiny\hspace{0.72cm} 3&\tiny\hspace{0.72cm} 4&\tiny\hspace{0.72cm} 5&\tiny\hspace{0.72cm} 6&\tiny\hspace{0.72cm} 7&\tiny \hspace{0.0cm}8\\
(g) $t=\frac{3\upi}{2}$&\multicolumn{8}{c}{\hspace{0.29cm}\vspace{-0.2cm}\includegraphics[scale=0.3]{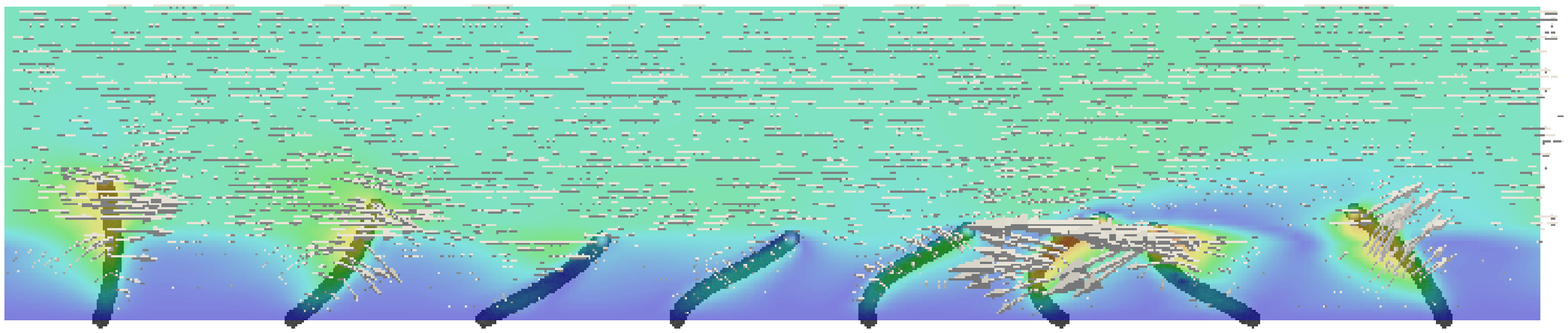}}\\
&\tiny\hspace{0.75cm} 1&\tiny\hspace{0.72cm} 2&\tiny\hspace{0.72cm} 3&\tiny\hspace{0.72cm} 4&\tiny\hspace{0.72cm} 5&\tiny\hspace{0.72cm} 6&\tiny\hspace{0.72cm} 7&\tiny \hspace{0.0cm}8\\
(h) $t=\frac{7\upi}{4}$&\multicolumn{8}{c}{\hspace{0.29cm}\vspace{-0.2cm}\includegraphics[scale=0.3]{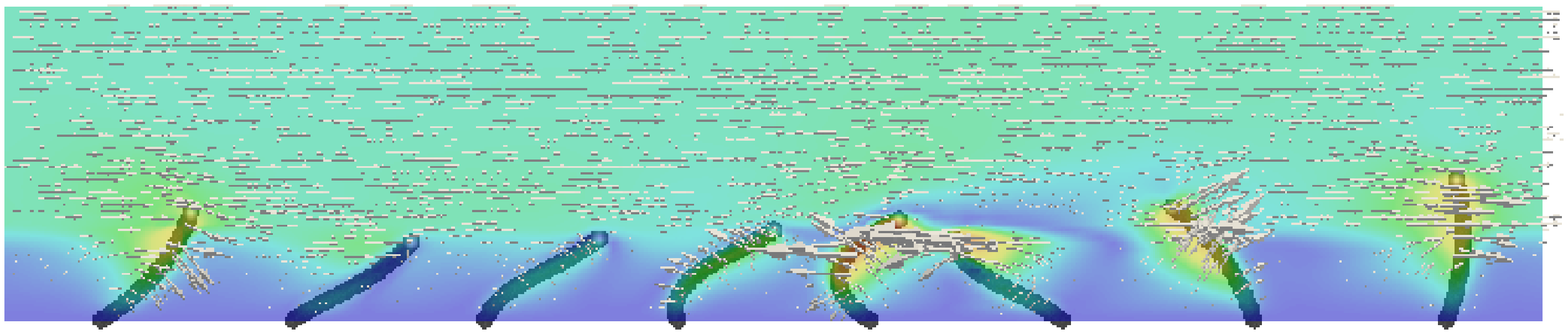}}\\
&\tiny\hspace{0.75cm} 1&\tiny\hspace{0.72cm} 2&\tiny\hspace{0.72cm} 3&\tiny\hspace{0.72cm} 4&\tiny\hspace{0.72cm} 5&\tiny\hspace{0.72cm} 6&\tiny\hspace{0.72cm} 7&\tiny \hspace{0.0cm}8\\
\multicolumn{9}{c}{\vspace{-0.2cm}\includegraphics[scale=0.3]{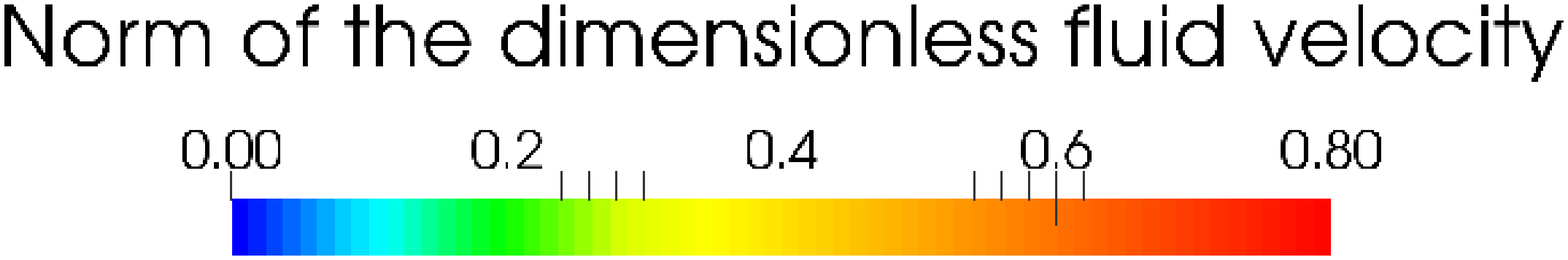}}

\end{tabular}
\caption{Snapshots of the fluid velocity taken at $8$ different instants during the beating cycle of the cilium located at the left of the images. The cilia are beating in an antipleptic motion with a phase lag $\Delta \Phi=\upi/4$, and a spacing of $a/L=2$. (a) $t=0$: the recovery phase of the first cilium begins; (b-e) recovery phase; (f-h) stroke phase. Once (h) is completed, a new cycle begins: (h) $\rightarrow$ (a). The figures show contours of the norm of the normalized fluid velocity $\|\vec{V}_{f}\|$ such that: $\vec{V}_{f}=\frac{\vec{V}_{f}^{dim}}{V_{f}^{ref}}$ with $V^{ref}=\lambda/T$ the reference speed of the present system.}
\label{Figure9}

\end{figure}

To interpret the evolution of $\epsilon_{PN}$ in figure \ref{Figure8}, it is necessary to examine the topology of the flow during this beating cycle, shown in figure \ref{Figure9}. As cilium 1 begins its recovery phase (case (a) of figure \ref{Figure9}), there is no reversal flow at the periodic boundary frontier. Hence, $\epsilon_{PN}=1$, as it is visible in figure \ref{Figure8} at the point (a). Then, when cilium 1 carries on its recovery phase (case (b) of figure \ref{Figure9}), no reversal flow can be observed since cilium 1 is still at the beginning of its recovery phase and cilium 8 is finishing its stroke phase. At $t=\upi/2$ (case (c) of figure \ref{Figure9}), both cilia 1 and 8 are in their early recovery phase and the directional pushing efficiency begins to decrease (see points (b) and (c) on figure \ref{Figure8}). At $t=3\upi/4$ (case (d) of figure \ref{Figure9}), cilium 1 is reaching the maximal speed of its recovery phase. The negative flow generated is important, and the efficiency quickly drops: see point (d) on figure \ref{Figure8}. Between $t=3\upi/4$ and $t=\upi$, the directional pushing efficiency weakly increases. This is due to the positive velocity near the base of cilium 1 which is doing a ``whip-like'' motion: while its free end is still going to the left, its base starts moving on the right. Case (e) of figure \ref{Figure9} corresponds to the end of the recovery phase of the first cilium and at the same moment, cilium 8 is at its maximal negative velocity: henceforth the directional pushing efficiency is still decreasing. At $t=5\upi/4$ (case (f) of figure \ref{Figure9}), cilium 8 has finished its recovery phase and cilium 1 is in the middle of its stroke phase. It counteracts the reverse flow created earlier and $\epsilon_{PN}$ finally increases. At $t=3\upi/2$ and $t=7\upi/4$ (cases (g) and (h) of figure \ref{Figure9}), both cilia 1 and 8 are in their stroke phases, and the flow generated is purely in the positive $x$-direction: see points (g) and (h) on figure \ref{Figure8}.

\begin{figure}
\centering
\begin{tabular}{cc}

\includegraphics[scale=0.25]{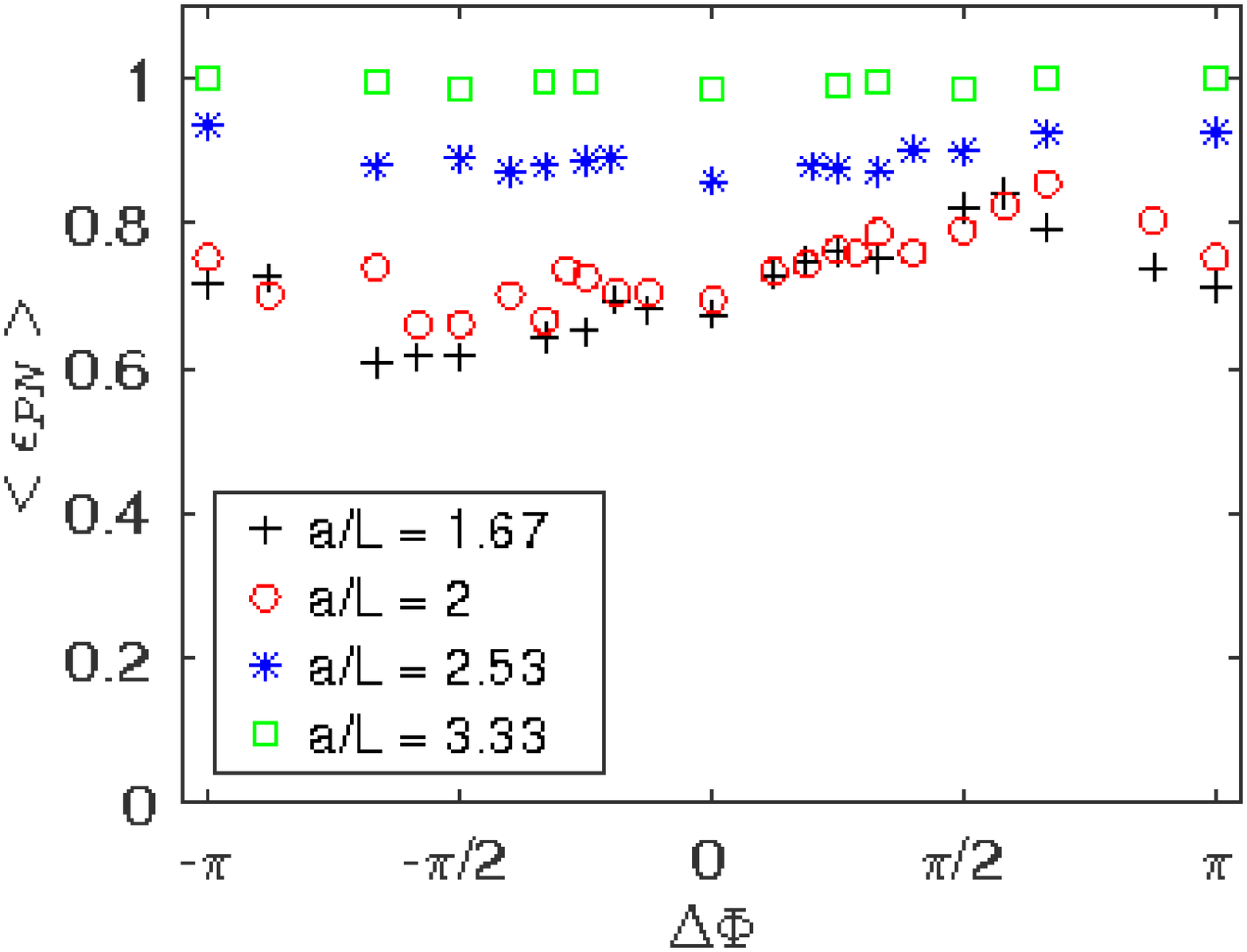} &
\includegraphics[scale=0.25]{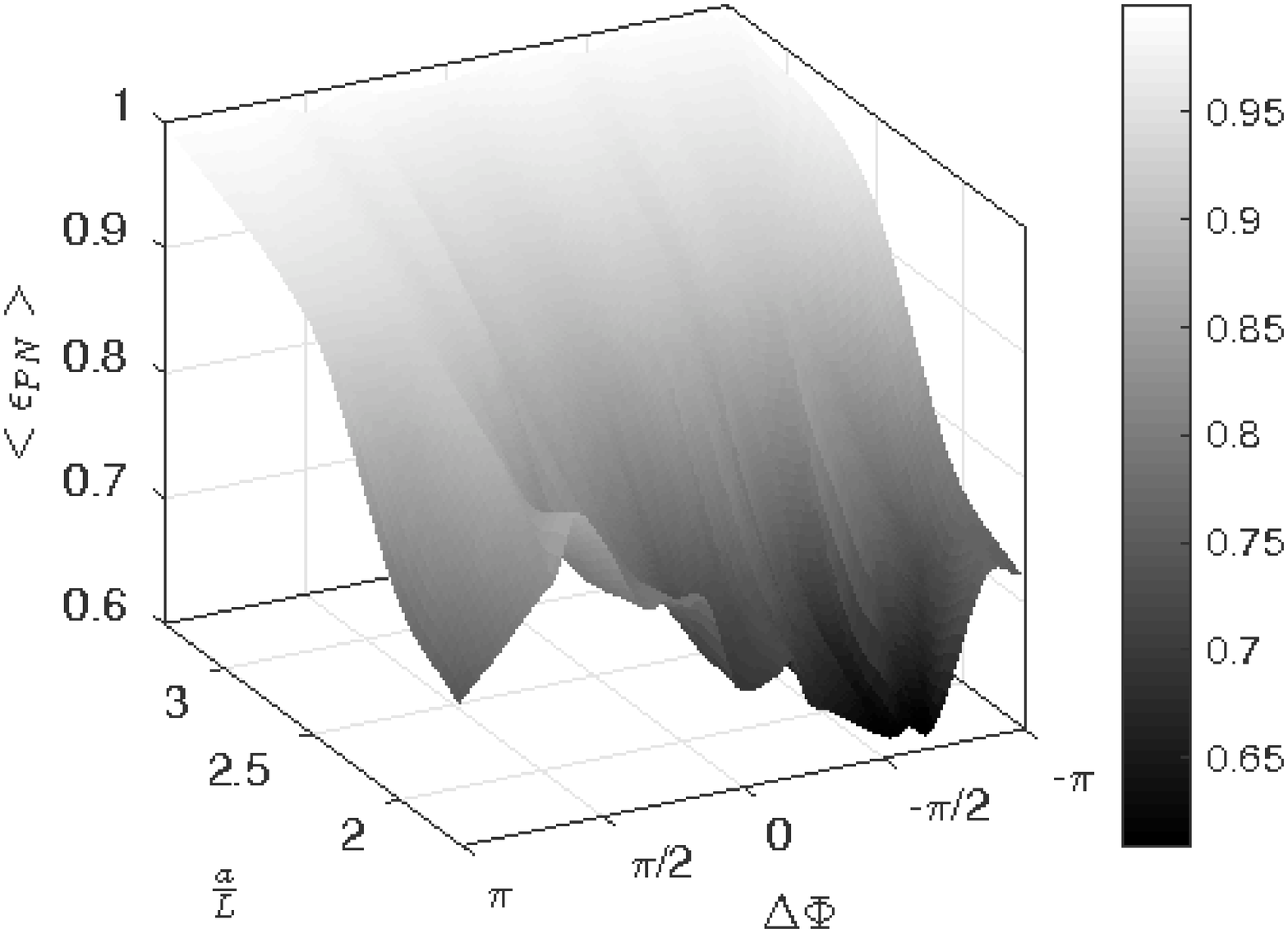} \\
(a)&(b)\\
\end{tabular}
\caption{Mean directional pushing efficiency $<\epsilon_{PN}>$ over a beating cycle as a function of the phase lag $\Delta \Phi$ for different cilia spacing $a/L$.}
\label{Figure10}
\end{figure}

It is now interesting to compare the average directional pushing efficiency $<\epsilon_{PN}>$ of the different kinds of synchronization over a full beating cycle (see figures \ref{Figure10} (a) and (b)). Before going further into details, let us recall the fact that the current efficiency $<\epsilon_{PN}>$ is a criterium qualifying the non-isotropy of the transport. Thus, it only gives an insight of the capacity of the cilia to transport the flow in an unidirectional way, but does not give any quantitative information on the volume of fluid flow, which is effectively displaced. 
Figures \ref{Figure10} (a) and (b) show an interesting phenomenon: systems with larger cilia spacing have a better capacity to transport the flow in the same direction. Another intriguing fact is that, for the smallest cilia spacing ($a/L=1.67$, figure \ref{Figure10}), antipleptic MCW seem to have a better ability to create an unidirectional flow compared to synchronized or symplectic motion, with a peak for the average efficiency around $\upi/2$. It agrees particularly well with the results of \citet{Gauger2009} who reported that antipleptic MCW were more efficient than the synchronized motion of cilia, itself being more efficient that the symplectic case. It also partially agrees with the results of \citet{Ding} who obtained peaks of efficiency for phase lags around $\pm \upi/2$, with a stronger one for the antipleptic case. In the present study, the minimal efficiency for metachronal motion is reached for a symplectic MCW with a phase lag of approximately $-\upi/2$.

When the cilia spacing is increased to $a/L=2$ (figure \ref{Figure10}), a peak of efficiency appears for the antipleptic MCW for $\Delta \Phi=2\upi/3$; and the lower values are reached in the symplectic case for $\Delta \Phi \approx -\upi/2$ and $\Delta \Phi \approx -\upi/3$. These trends will be found again in the results shown in \S \ref{ssec:Compared_hydro_eff}.

Then, if the cilia spacing is increased above a limit value ($a/L=3.33$, figure \ref{Figure10}), the directional pushing efficiency becomes equal to 1 for all cases (antipleptic, symplectic, and synchronized). It agrees well with the results obtained by \citet{Khaderi11} who concluded that ``the amount of flow enhancement depends on the inter-cilia spacing [but] the efficiency is not significantly influenced''. Indeed, as the cilia are set away from each other, the influence of their neighbours become negligible. 
Nevertheless, the authors reported a very low efficiency for synchronized cilia, whereas the present results always show a positive flow in the mucus phase for synchronized motions. As mentioned in \S \ref{ssec:geometrical}, this is a direct consequence of the inertial effects ($\Rey>1$) in the present simulations. It is recalled that they only affect the synchronized case. It has carefully been checked that all other conclusions drawn for antipleptic and symplectic metachrony remain the same for $\Rey <1$ and $\Rey =20$. For a detailed study of the inertial effects at $\Rey =20$, see \S \ref{sssec:Reynolds}.
The present results are different from the results of \citet{Gueron97}, as it appears that even when cilia spacing is higher than 2 cilium lengths, the influence of neighbouring cilia cannot be neglected (see the case $a/L=2.53$ for example). 

It is important to remember that this efficiency does not take into account the actual net flow volume transported. 

\subsubsection{Compared hydrodynamic efficiency}\label{ssec:Compared_hydro_eff}

After having investigated the capacity to transport the flow in the desired direction, it is now interesting to look at the actual flow volume displaced to see if a better efficiency to transport the flow directionally corresponds to a better flow transport. 
To do so, the global volumetric flow rate $Q_{v}$ over a unit volume of size ($1 \times 1 \times N_{z}$) is introduced:
\begin{align}
Q_{v}=N_{z}\frac{U^{*}dx^{2}}{L^{2}}
\end{align}
where $dx=1$ using the classical LBM normalization, and $U^{*}=U^{av}/U^{ref}$, with $U^{ref}$$=$$(\lambda/N_{cil})/T$ being the reference velocity and $U^{av}$$=$$\frac{1}{n_{i}n_{j}n_{k}}\sum_{i,j,k}U_{ijk}$ the average fluid velocity inside the whole domain.

\begin{figure}
\centering
\includegraphics[scale=0.4]{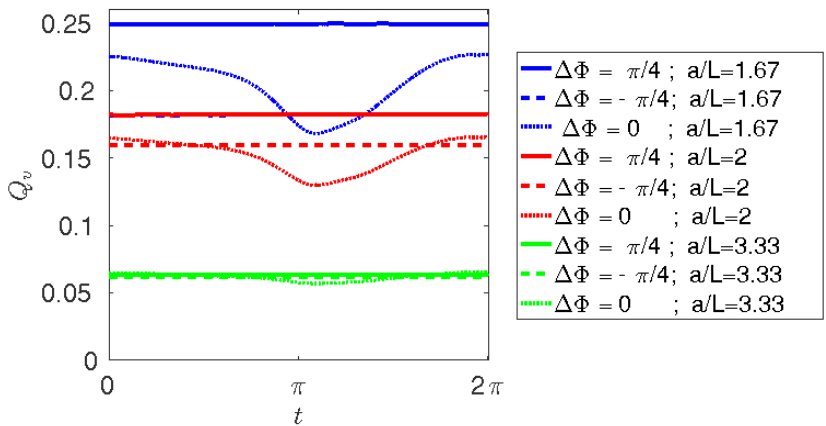}
\caption{Normalized volumetric flow rate $Q_{v}$ as a function of time over a beating cycle for different phase lags $\Delta \Phi$ and different cilia spacings $a/L$. The recovery phase occurs for $t \in [0,\upi]$ and the stroke phase for $t \in [\upi,2\upi]$.}
\label{Figure11}
\end{figure}

On figure \ref{Figure11}, the dimensionless volumetric flow rate $Q_{v}$ is plotted over one beating cycle for arrays of cilia with different phase lags ($\Delta \Phi=0$, $\Delta \Phi=-\upi/4$, and $\Delta \Phi= \upi/4$) and different cilia spacings. For cilia spacings equal to $a/L=1.67$ and $a/L=2$, the antipleptic MCW is the most efficient for transporting fluids. 
However, note that as the cilia spacing is increased, 
the ability of the antipleptic wave to transport fluids exhibits a huge decrease compared to the symplectic wave. 
For a larger cilia spacing ($a/L=3.33$), the fluid transport is not impacted anymore by metachrony. Finally, no flow reversal occurs for the synchronized cases: this is the consequence of working at $\Rey=20$.

\begin{figure}
\centering
\begin{tabular}{cc}

\includegraphics[scale=0.25]{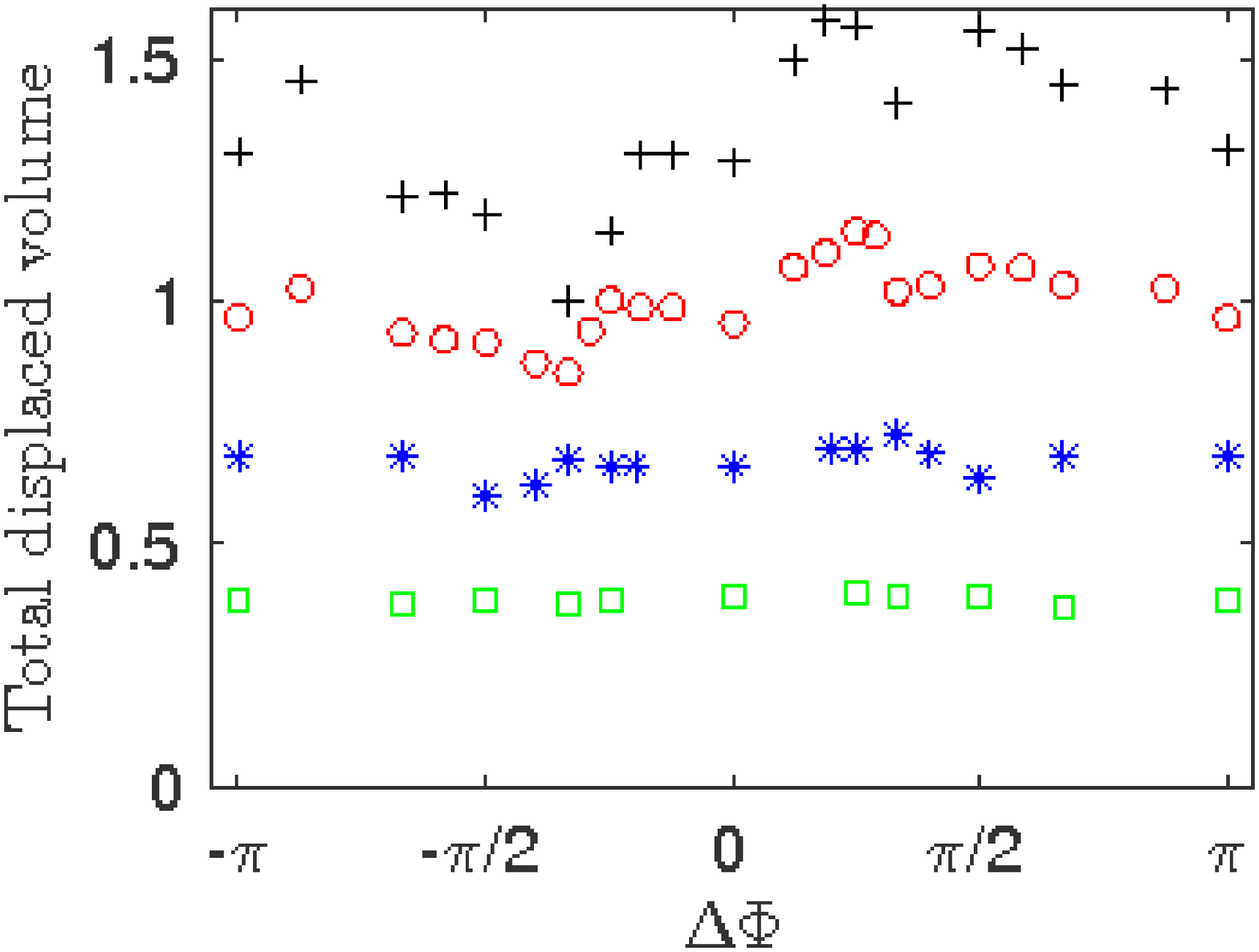} &
\includegraphics[scale=0.25]{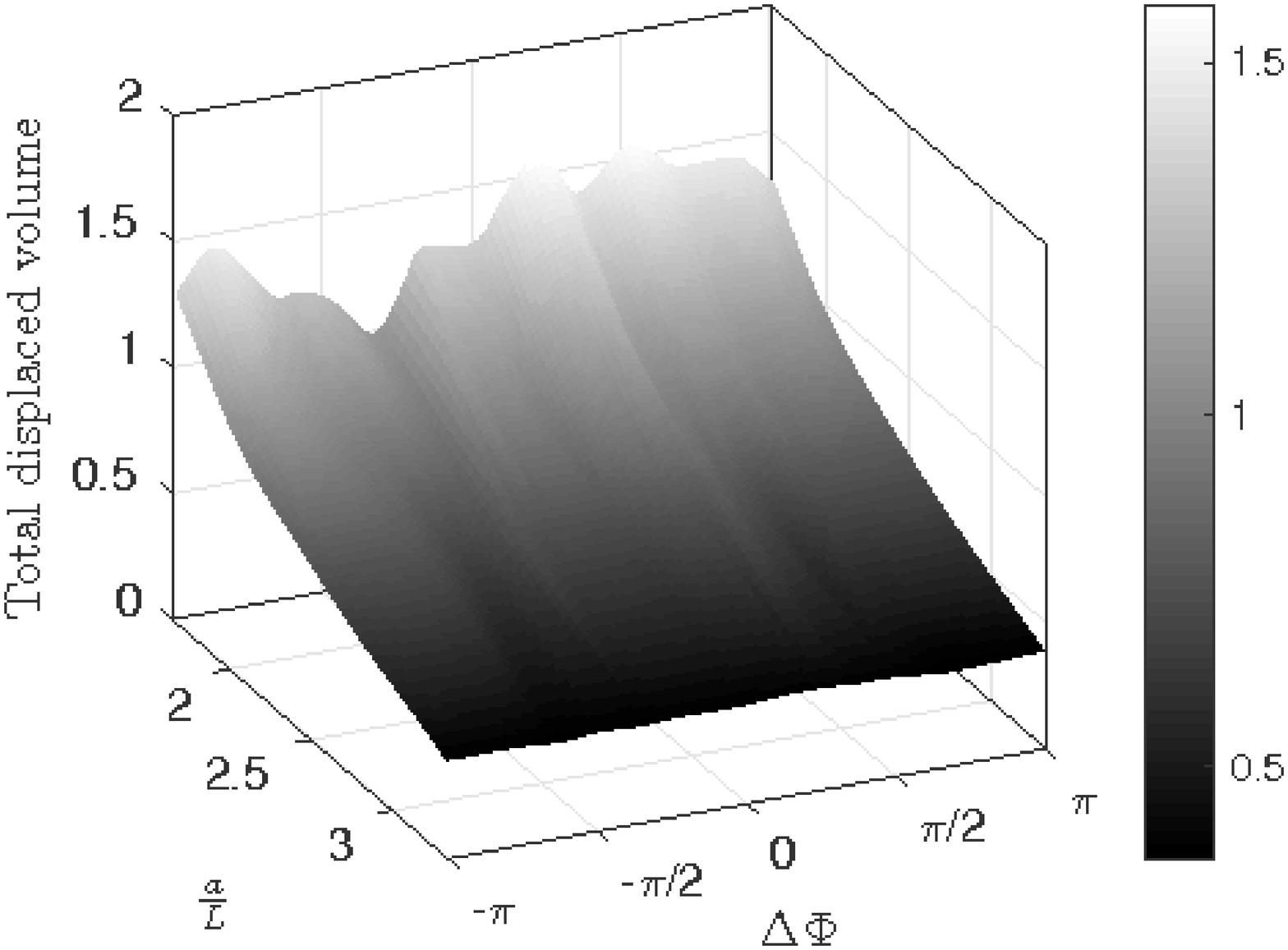}\\
(a)&(b)\\
\end{tabular}
\caption{Total dimensionless displaced flow volume generated by an array of cilia over a beating cycle for different phase lags and cilia spacings. $+$: $a/L=1.67$; $\textcolor{red}{\bigcirc}$: $a/L=2$; $\textcolor{blue}{\ast}$: $a/L=2.53$; $\textcolor{green}{\square}$: $a/L=3.33$.}
\label{Figure12}
\end{figure}

The total volume of fluid displaced during a beating cycle for the different phase lags is compared on figures \ref{Figure12} (a) and (b). For a small cilia spacing ($a/L=1.67$), the efficiency of the antipleptic metachrony is obvious. It agrees well with the results of \citet{Khaderi11} who observed a larger net flow produced by antipleptic metachrony for this value of cilia spacing. Symplectic waves appear to be less or at best equally efficient than antipleptic motion, except for $\Delta \Phi = - 7\upi/8$ for $a/L=1.67$ where there is a peak in the total displaced volume of flow. On the contrary, negative peaks are found for $a/L=1.67$ and $a/L=2$ for $\Delta \Phi = -\upi/3$. There are two neighbouring maxima at $\Delta \Phi = \upi/4$ and $\Delta \Phi = \upi/2$ for $a/L=1.67$ and $a/L=2$ respectively, indicating that specific phase lags are more able to generate a strong flow. At this point, it is worth remembering that in the present model, the recovery and stroke phases occur in the same plane which is not the case in real configurations. Accordingly to \citet{Downton}, who studied both 2D and 3D beating patterns, it is expected that in presence of 3D beating patterns, both the directional pushing efficiency $\epsilon_{PN}$ and total displaced volume of flow would increase. It is also expected that a small flow might occur in the $y$-direction.

\subsubsection{Energetic cost of the metachronal waves}\label{ssec:NumPower}
The previous sections have been dedicated to the study of the directional transport and the volume displacements of fluids. An energetic perspective is now introduced, to characterize completely the potential benefits of metachrony.

The average power spent by the cilia during a beating cycle is given by:
\begin{equation}
P_{cil}=\frac{\sum_{s,i}\boldsymbol{V}_{i}^{s}\cdot(\boldsymbol{F}_m^{i}+\boldsymbol{F}_{PCL}^{i})}{N_{cilia}}
\end{equation}
using the forces illustrated in figure \ref{Figure2}. The power spent is averaged over several beating cycles. To have a dimensionless power $P^{*}$, the power $P^{\infty}$ spent by an isolated cilium ($a/L=10$) is computed such that $P^{*}$$=$$P_{cil}/P^{\infty}$.

\begin{figure}
\centering
\begin{tabular}{cc}

\includegraphics[scale=0.25]{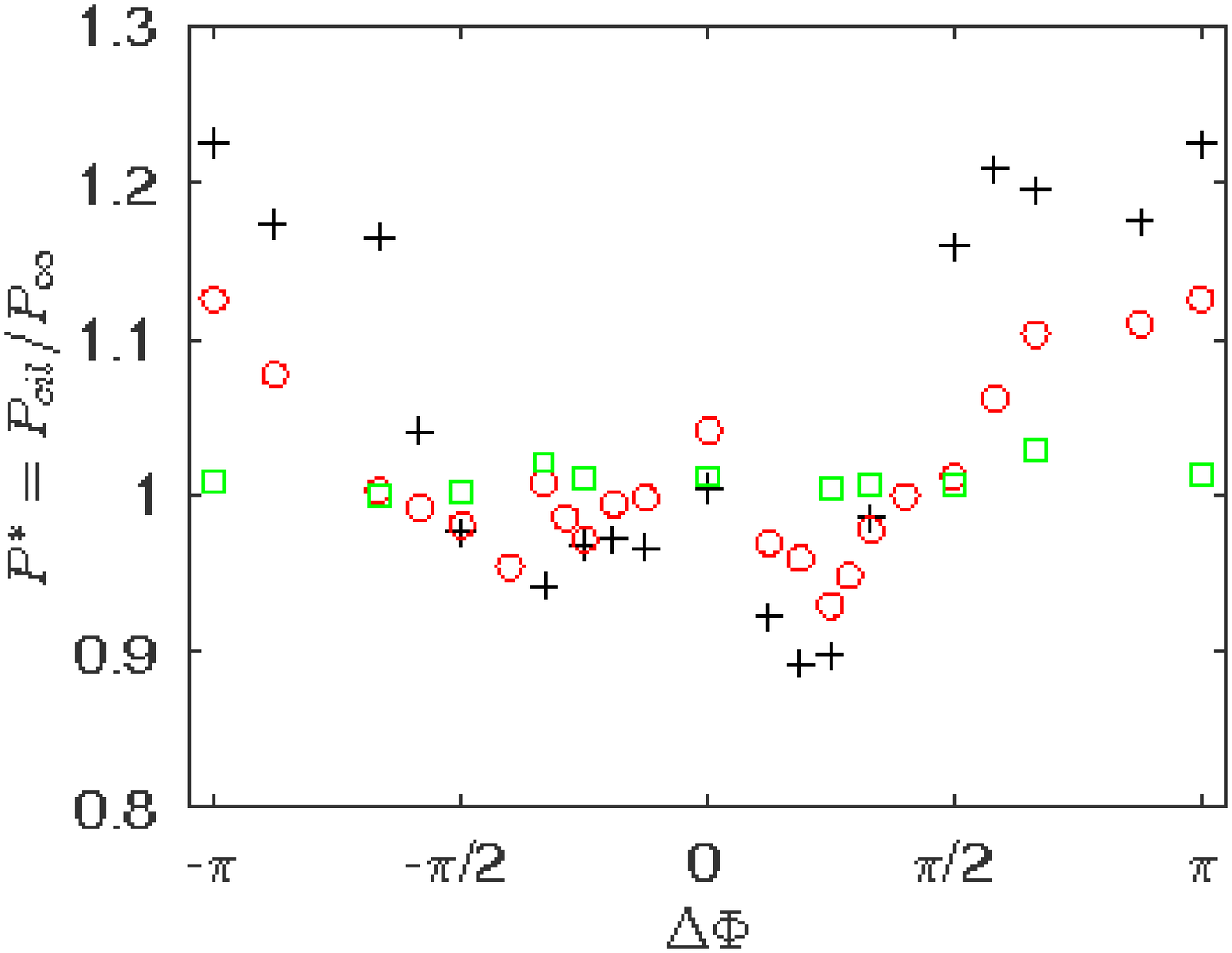} &
\includegraphics[scale=0.25]{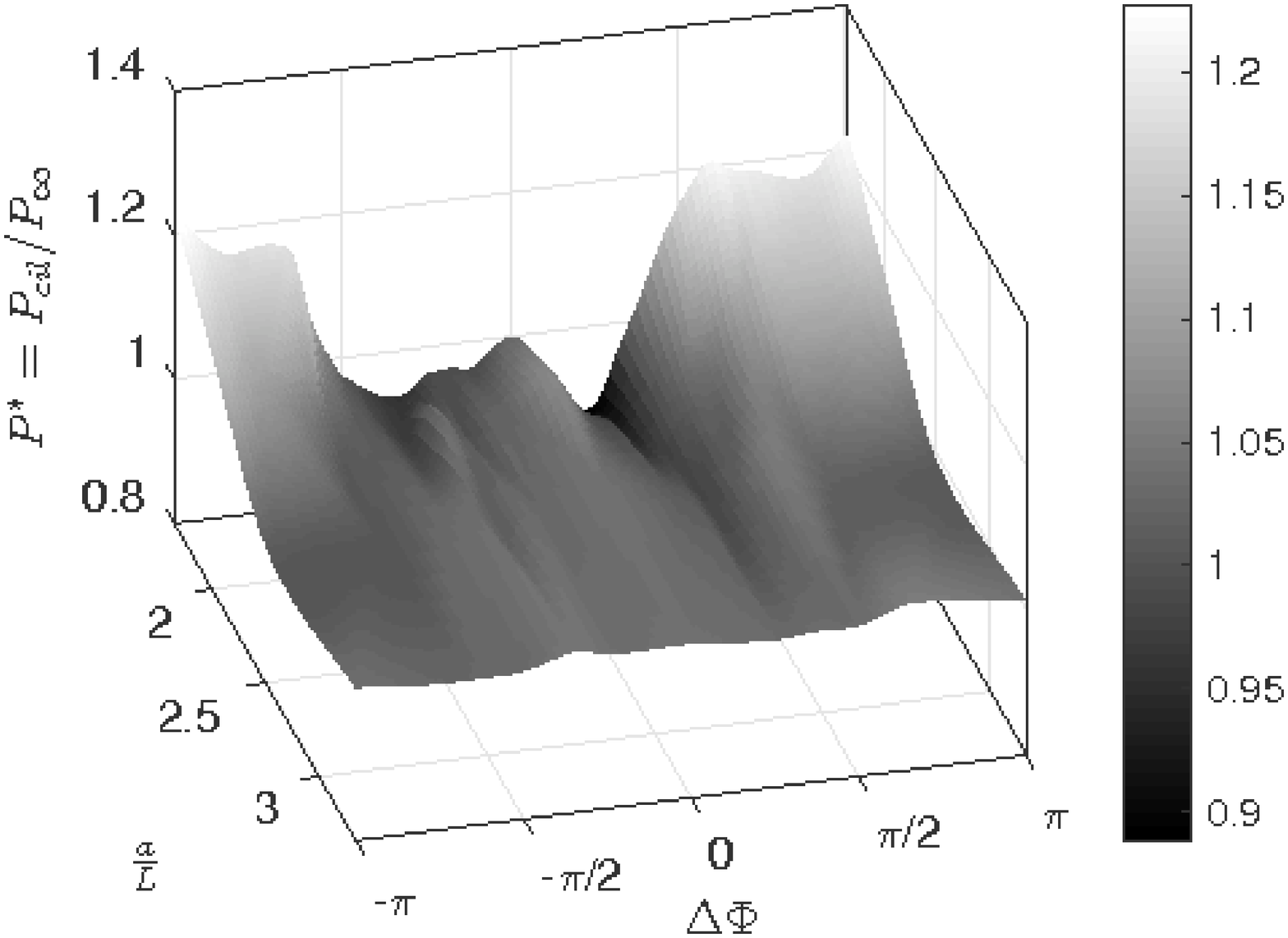} \\
(a)&(b)\\

\end{tabular}
\caption{Power spent by the system for different phase lags $\Delta \Phi$ and different cilia spacings. $+$: $a/L=1.67$; $\textcolor{red}{\bigcirc}$: $a/L=2$; $\textcolor{green}{\square}$: $a/L=3.33$}
\label{Figure13}
\end{figure}

Before going any further, it is important to remember here that the only parameter that differs between each value of phase lag $\Delta \Phi$ for a given cilia spacing $a/L$, is the size of the domain $N_{x}$ over the $x$-direction, hence the number of cilia acting in one wavelength. Moreover, between different cilia spacings $a/L$, the space $a$ between two cilia is modified in both the $x$ and $y$-directions, and the sizes $N_{x}$ and $N_{y}$ of the domain must be changed accordingly. As a consequence, if the cilia spacing is increased, the density of cilia in both the $x$ and $y$-directions is decreased. In all the simulations presented in \S \ref{sec:Study}, all the other parameters (ratio of viscosity $r_{\nu}$, beating frequency $f$, length of the cilia $L$, etc.) are fixed.

Figures \ref{Figure13} (a) and (b) show the average dimensionless power as a function of $\Delta \Phi$. For small values of phase lags ($| \Delta \Phi | \le \upi/2$), the average power spent decreases to a minimal value for $\Delta \Phi \approx \upi/4$. For this particular value, the system spends less power than the synchronized case: the antipleptic MCW allows the system composed of all cilia to encounter a smaller resistance from the surrounding fluids.

Nevertheless, it is obvious, from a fluid mechanics point of view, that when cilia beat together in a synchronized way, the viscous resistance felt by each cilium is reduced. 

In the case of metachrony, the behaviour is dual. Indeed, cilia in the stroke phase of the antipleptic motion encounter a stronger viscous resistance from the flow due to their added respective remoteness. Therefore, compared to the synchronized motion, they will transfer more energy to the flow; and the energy transferred will be entirely used to propel the fluids in the desired direction. The opposite is true during the recovery stroke of the antipleptic motion: the clustered behaviour of the cilia allows them to experience a lower viscous resistance from the flow, and therefore to limit the amount of power transferred during this phase. But, globally, the system with antipleptic MCW and $|\Delta \Phi| < \upi/2$ encounter less viscous resistance than the synchronized motion of cilia, as it can be seen on figures \ref{Figure13} (a) and (b). It results in a better efficiency of the cilia with antipleptic metachronal motion to transfer their momentum to the flow, while in the meantime requiring less power. For the symplectic motion, a similar phenomenon happens: cilia in their stroke phase are clustered and therefore unable to fully exert their pushing action, while cilia in their recovery phase are away from each other and generate a stronger reversal flow compared to the antipleptic motion. It results in a lower capacity of the symplectic MCW to transport mucus, while still allowing the system to spend less power than synchronously beating cilia for $|\Delta \Phi| < \upi/2$. An energetic ratio will be introduced in \S \ref{ssec:displacement_ration} to quantify, for such value of $\Delta \Phi$, the capacity of the cilia with antipleptic or symplectic metachrony to transfer their momentum to the flow.

\begin{figure}
\centering
\begin{tabular}{cc}
\includegraphics[scale=0.23]{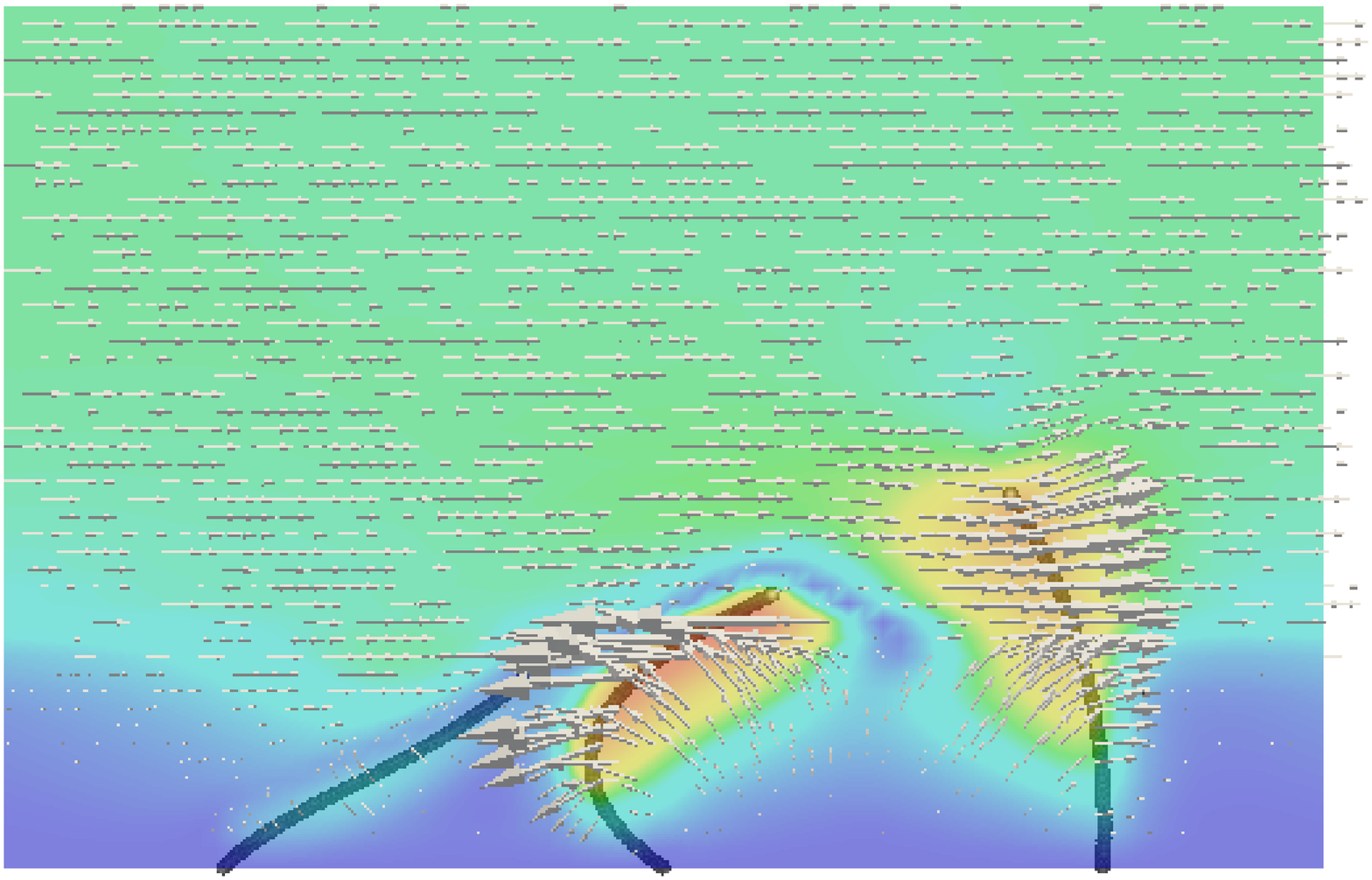}& \includegraphics[scale=0.23]{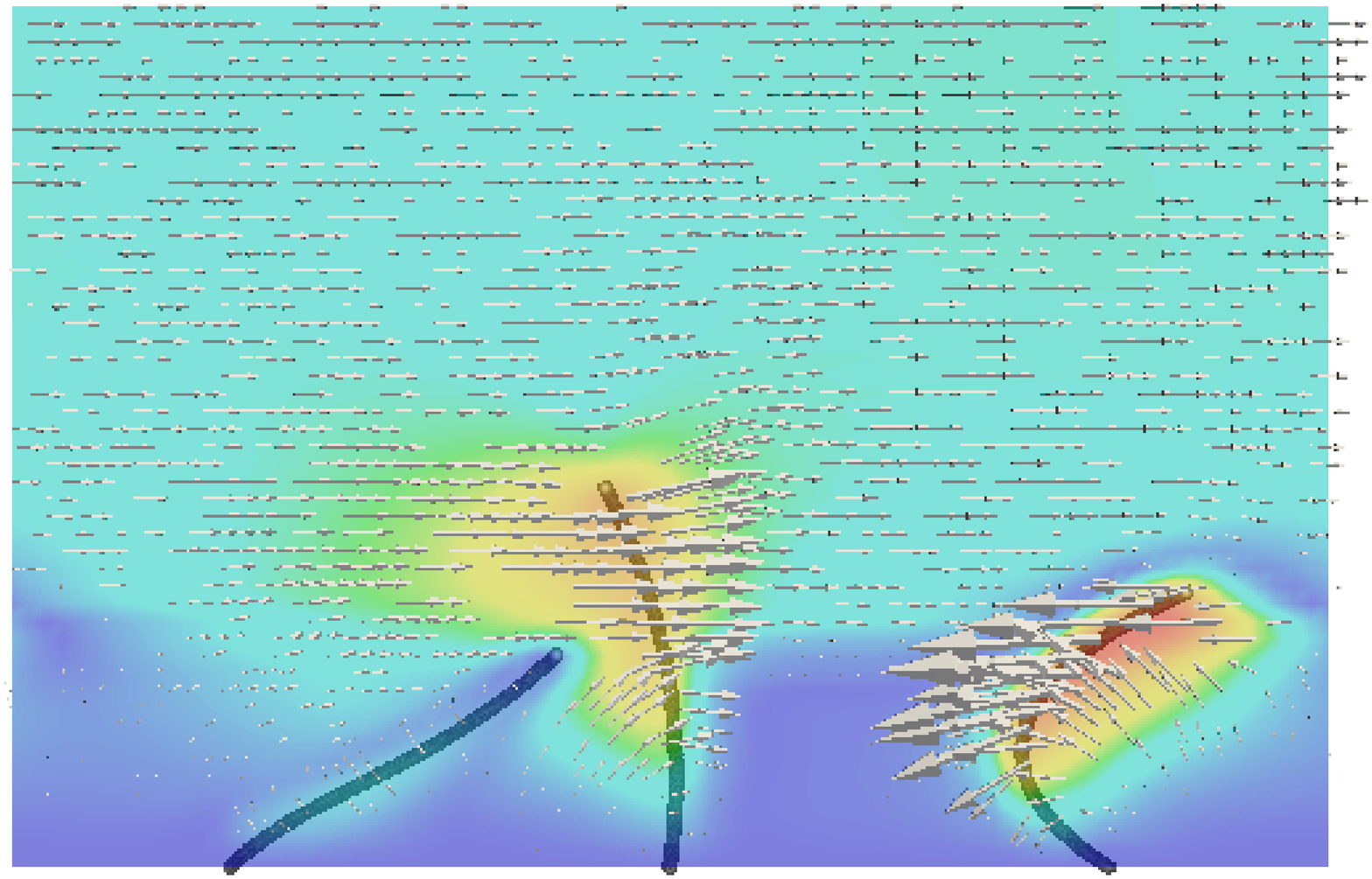}\\
  (a) &  (b) \\
\multicolumn{2}{c}{\includegraphics[scale=0.2]{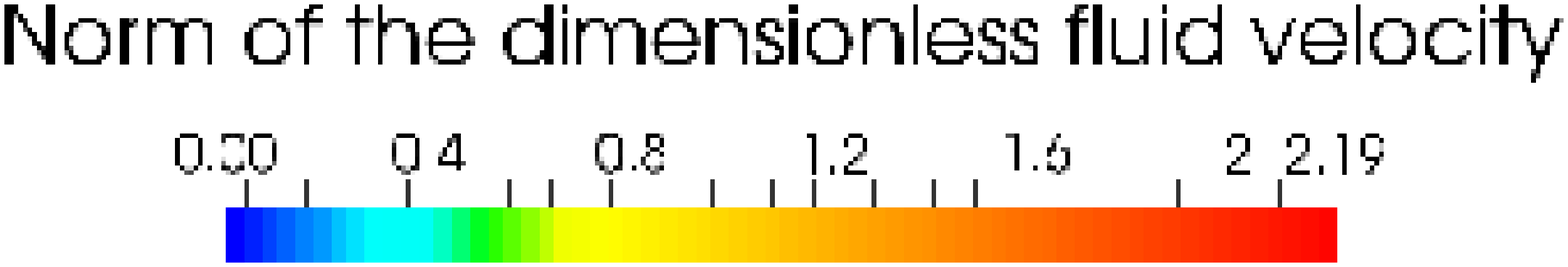} } \\

\end{tabular}
\caption{Comparison of the flow generated by an antipleptic and a symplectic wave for a the cilia spacing $a/L=1.67$. (a): Antipleptic MCW with $\Delta \Phi=2\upi/3$. (b): Symplectic MCW with $\Delta \Phi=-2\upi/3$. The plane is colored with the magnitude of the dimensionless fluid velocity.}
\label{Figure14}
\end{figure}

This is not true for large phase lags ($| \Delta \Phi | \ge \upi/2$): the system with antipleptic or symplectic metachrony spends more power than the synchronous case. One can suppose that it is a direct consequence of the reduced number of cilia in a wave length. Then, to explain the better efficiency of the antipleptic MCW over the symplectic ones for large phase lags, an investigation of the flow topology is necessary. Figure \ref{Figure14} shows an antipleptic MCW (on the left) and a symplectic MCW (on the right) at the same time for a phase lag $\Delta \Phi =\pm 2\upi/3$. For such value of $\Delta \Phi$, the antipleptic wave is more efficient for transporting the mucus (see figures \ref{Figure12} (a) and (b)), although the average power spent by both system is relatively similar (see figures \ref{Figure13} (a) and (b)). One can easily see the main difference between both cases: in the symplectic case, the cilium in stroke phase encounters the reversal flow generated by the other cilium in recovery phase. When these two secondary flows meet, vortices are generated and the global transport of fluid is less efficient, as a fraction of the energy transferred to the flow is used to cancel this reversal flow. On the contrary, in the antipleptic case, the cilium in stroke phase does not immediately feel the influence of the reversal flow created by the other cilium in recovery phase, which is behind him. Henceforth, this cilium is able to fully exert its pushing effect on the mucus phase. Moreover, for the antipleptic case, there is a suction effect due to the combined motion of the cilia in the stroke and recovery phases, maximizing the propulsion of mucus. One can then expect a weak blowing effect between the cilia in the stroke and recovery phases of symplectic motion.

%

When the cilia spacing is large (see the case $a/L=3.33$ of figures \ref{Figure13} (a) and (b) for example), the power spent by the system is the same for all phase lags: the cilia are too far from each other to be impacted by the phase difference of their immediate neighbours. One can observe that for $\Delta \Phi=0$ (i.e. for synchronized beating), the average power spent by a cilium $P^{*}$ is equal to 1 in figure \ref{Figure13} (a) for the cilia spacings $a/L=1.67$ and $a/L=3.33$, meaning that synchronous systems spend the same amount of power as an isolated cilium. It shows the beneficial cost of antipleptic metachrony for small phase lags.

\subsubsection{Displacement ratio}\label{ssec:displacement_ration}
Inspired by the work of \citet{KimNetz}, a displacement ratio, which can be seen as the transport efficiency of the waves, is introduced to quantify the capacity of a given system to transport particles, with respect to a given amount of power. 
In that context, $\eta_{1}$ is defined by the mean displacement over the $x$-direction during one beating cycle, divided by the mean power that a cilium had to spend during this beating cycle. Since the main purpose of mucociliary clearance is to transport mucus, and since experimental data \citep{Winters} report that the total thickness in the vertical direction of the mucus layer is in the range $[1.4L;10L]$, values for the displacement were taken on an arbitrary plane $z/L=3.2$ near the extremity of the domain. 
To obtain a value for the displacement, the instantaneous average fluid velocity over the $x$-direction is computed, and the resulting value is then multiplied by the period of a full beating cycle, giving the mean displacement $<d_{x}>$ over one period on the ($x$,$y$,$3.2L$) plane. 

By dividing this mean displacement with appropriate quantities, a dimensionless expression of the displacement ratio is obtained: 
\begin{align}
\eta_{1}=\displaystyle \frac{<d_{x}>\displaystyle \frac{N_{cil}}{\lambda}}{P^{*}}
\end{align}
In the synchronized case, i.e. $\Delta \Phi=0$, $\lambda$ is infinite and thus the size of the domain over the $x$ direction was used and divided by the number of cilia.

\begin{figure}
\centering
\begin{tabular}{cc}
\includegraphics[scale=0.25]{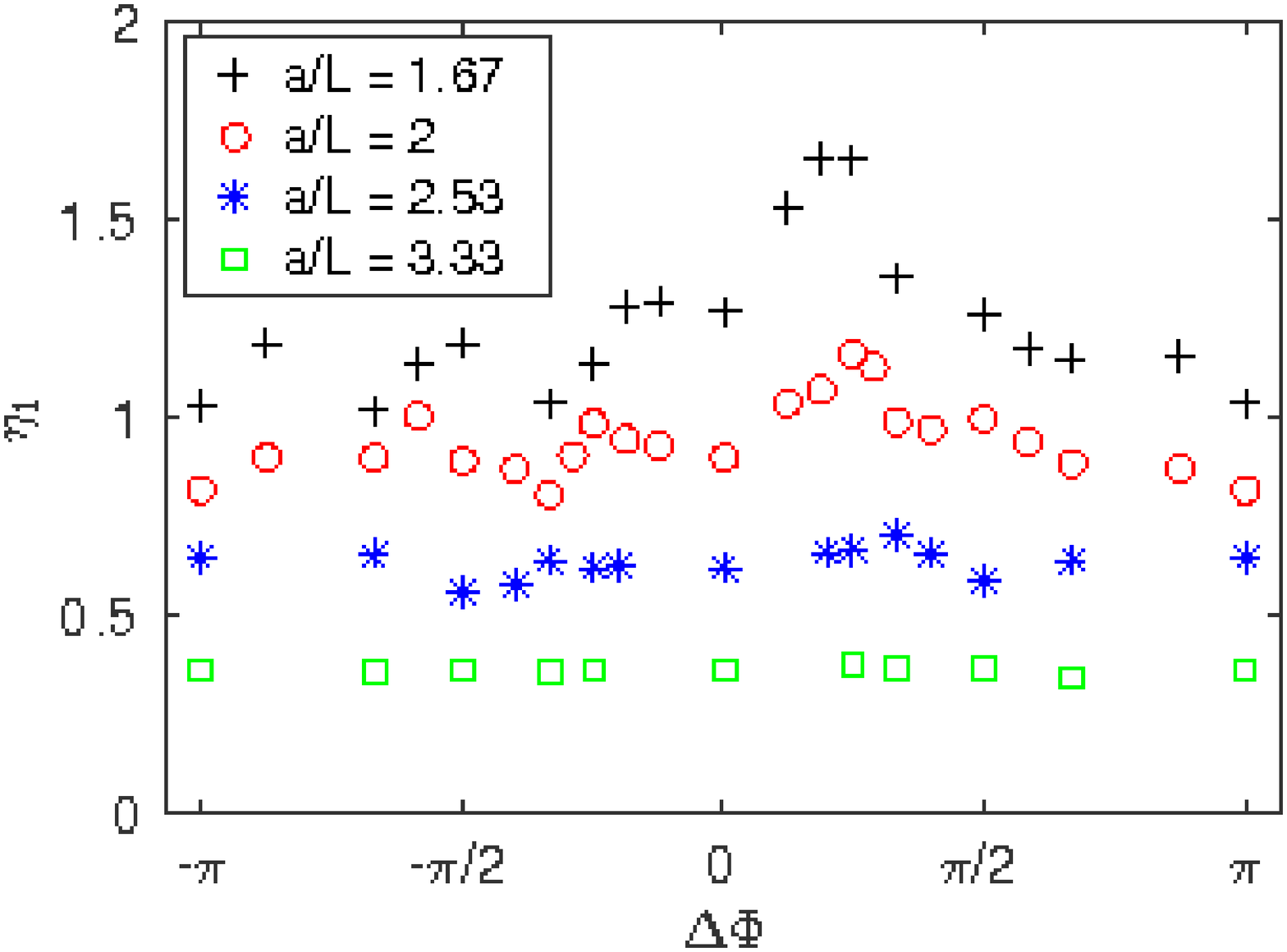} &
\includegraphics[scale=0.25]{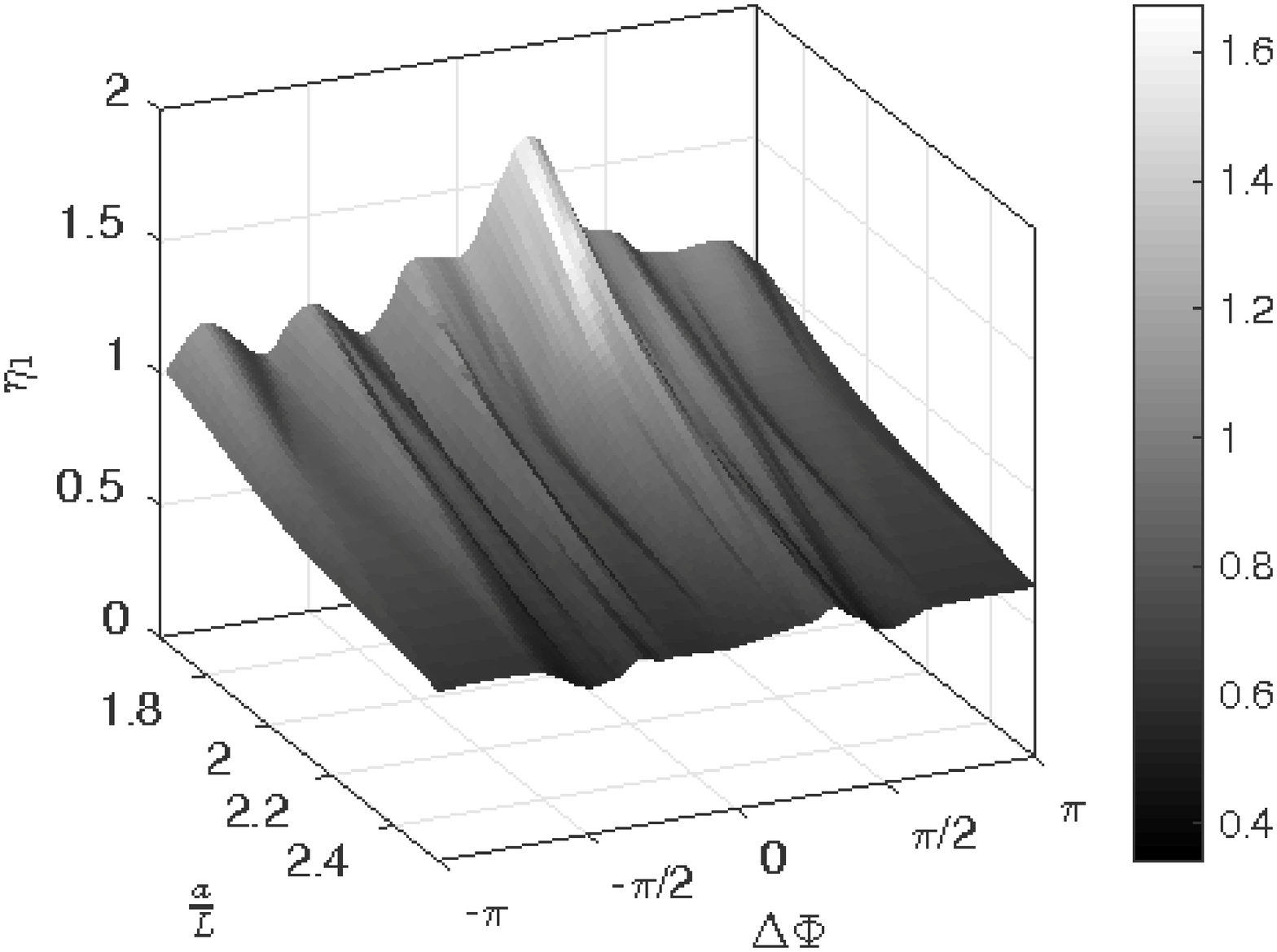} \\
(a)&(b)\\
\end{tabular}
\caption{Displacement ratio $\eta_{1}$ as a function of the phase lag $\Delta \Phi$ for different cilia spacings $a/L$.}
\label{Figure15}
\end{figure}

In figures \ref{Figure15} (a) and (b), one can see, from an energetic point of view, the superiority of the antipleptic wave to transport mucus for small cilia spacings, which confirms the previous findings. 
More than that, for the smallest cilia spacing ($a/L=1.67$), a clear peak of efficiency can be seen for antipleptic MCW with $\Delta \Phi=\upi/4$.
If the cilia spacing is increased ($a/L=2$), similar results are found. Nevertheless, for $a/L=2.53$, a different behaviour occurs: the displacement ratio is found to be the worst for $\Delta \Phi = \pm \upi/2$. Then, for cilia far away from each other, the displacement ratio $\eta_{1}$ remains constant for all phase lags (the cilia do not feel the influence of the others anymore). 

Similar results regarding the efficiency of the antipleptic MCW are found in \citet{Osterman}, where the optimal beating pattern is investigated. In their study, they 
observed that antipleptic MCW were often the most efficient. They also found, as it is the case in the present study, that increasing the density of cilia results in an increase of the efficiency up to a critical point where clustering becomes counter-productive. 

\subsubsection{Mixing}\label{sssec:mixing}
\begin{figure}
\centering
\begin{tabular}{cc}
\includegraphics[scale=0.25]{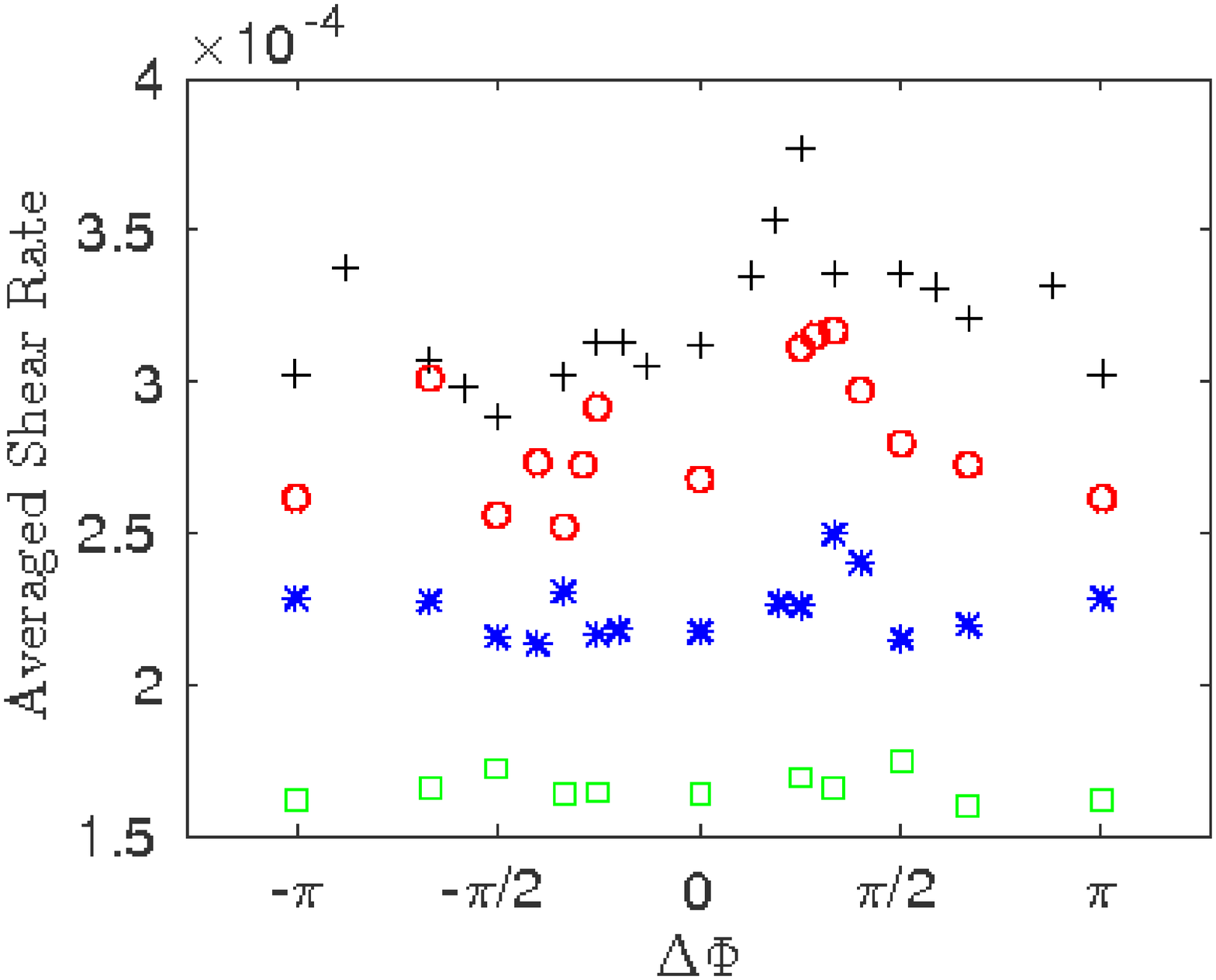} &
\includegraphics[scale=0.25]{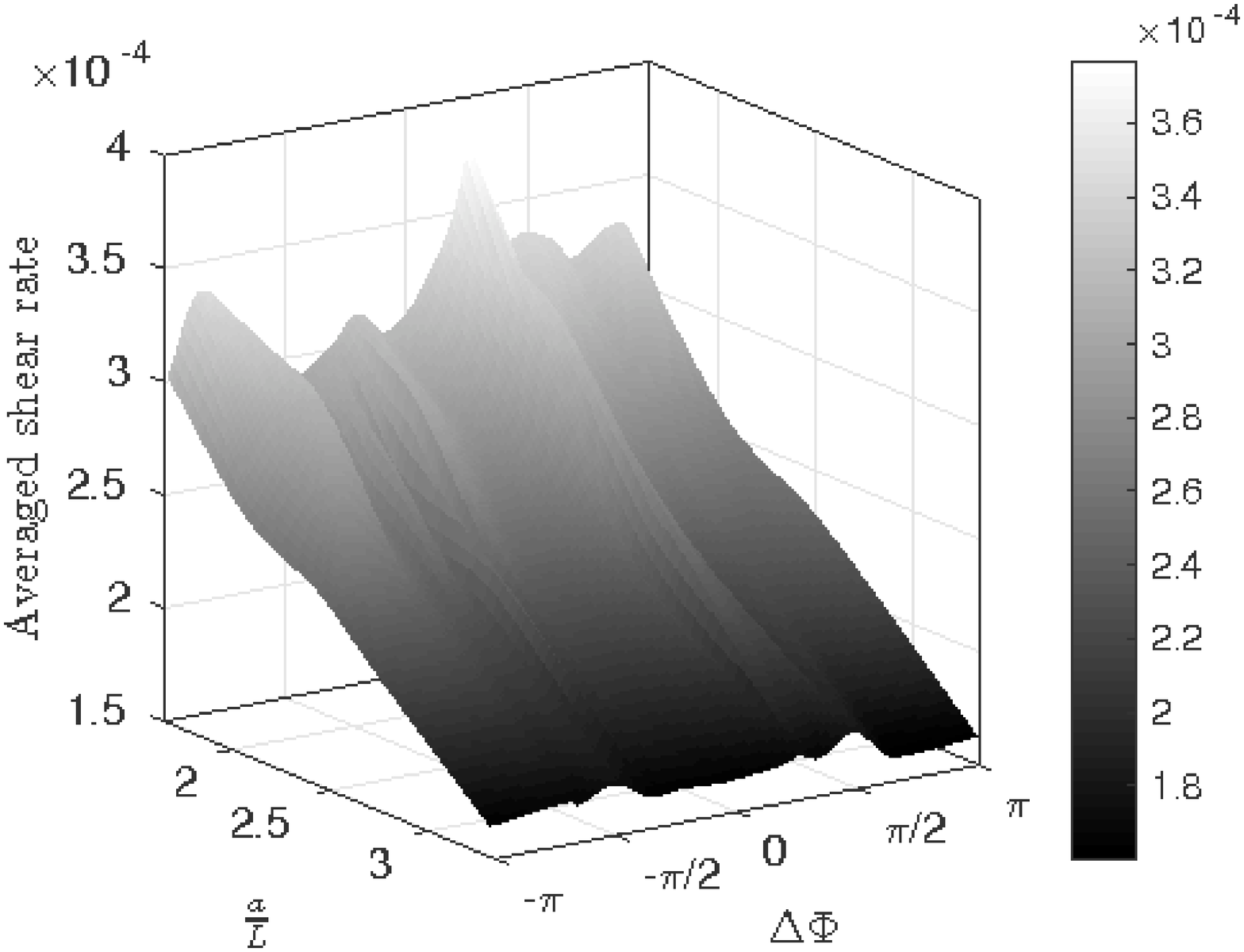} \\
(a) & (b)
\end{tabular}
\caption{Average stretching rate in the transport and mixing areas as a function of the phase lag $\Delta \Phi$ for different cilia spacings. (a) $+$: $a/L=1.67$; $\textcolor{red}{\bigcirc}$: $a/L=2$; $\textcolor{blue}{*}$: $a/L=2.53$; $\textcolor{green}{\square}$: $a/L=3.33$ (b) 3D view of the corresponding plot.}
\label{Figure16}
\end{figure}

To investigate how the mixing can be enhanced by metachrony, the average stretching rate over the transport and mixing areas during a beating period is computed. Figure \ref{Figure16} shows the results obtained for all cilia spacings. Clearly, the antipleptic MCW is the most efficient for stretching fluids, hence to mix fluids. Clear peaks are visible for antipleptic waves with $\Delta \Phi=\upi /4$ and $a/L=1.67$; with $\Delta \Phi\approx\upi /3$ and $a/L=2$; and with $\Delta \Phi\approx\upi /3$ and $a/L=2.53$. On the contrary, symplectic MCW are almost always less efficient for mixing than antipleptic MCW, except for $\Delta \Phi=-7\upi/8$ and $a/L=1.67$ where there is an enhancement in the mixing. 
This peak is also present for the opposite phase lag $\Delta \Phi=7\upi/8$ where it reaches approximately the same value. 
This is perfectly coherent with all previous results, and partially with the ones of \citet{Ding} who observed, for an unique layer of fluid with an uniform viscosity and a cilia spacing $a/L=1.67$, the existence of two peaks in the shear rate: a weak one for symplectic waves for $\Delta \Phi=-\upi/2$, and a stronger one for antipleptic waves for $\Delta \Phi=\upi/2$. The main difference here is that in the present study, the symplectic MCW with $\Delta \Phi\approx-\upi/2$ seem to have the worst capacity for mixing, while the antipleptic MCW reach their full mixing capacity for $\Delta \Phi=\upi/4$. The combined study of figures \ref{Figure13} and \ref{Figure16} gives a good insight of the mixing efficiency of the system.

\subsubsection{Quantification of inertial effects}\label{sssec:Reynolds}
\begin{figure}
\centering
\includegraphics[scale=0.25]{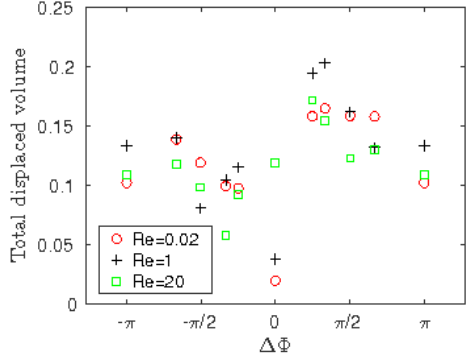}
\caption{Total displaced volume of fluid for $\Rey=0.02$, $\Rey=1$ and $\Rey=20$ as a function of the phase lag $\Delta \Phi$. The results of these simulations are obtained with $L=7$ lu and $a/L=1.67$.}
\label{Figure17}
\end{figure}

In this part, the influence of the Reynolds number is considered. 
The present objective is to compare the qualitative behaviour of the MCW between $\Rey<1$ and $\Rey>1$. To reach Reynolds numbers of the order of $10^{-2}$, the size of the cilia has been reduced to half the size of real cilia. The geometry is then divided by a factor 2 in the three spatial directions. The main consequence of such choice is that the computed quantities (fluxes, displaced volume of fluid, etc.) displayed here are approximately 8 times smaller compared to the results previously shown (as in figure \ref{Figure12}). However, the qualitative behaviour of the MCW remains the same, as well as the drag exerted by the cilia. Thus, a comparison of the MCW behaviour between $\Rey<1$ and $\Rey>1$ is possible. 
On Figure \ref{Figure17}, one can see the total displaced volume of fluid for $\Rey=0.02$, $\Rey=1$ and $\Rey=20$, and a cilia spacing $a/L=1.67$. A transition between $\Rey<1$ and $\Rey>1$ for the synchronous cases is expected, and can be seen on figure \ref{Figure17}. As mentioned in \S \ref{ssec:Compared_hydro_eff}, for $\Rey\le 1$, the synchronized cases are less efficient than the symplectic cases. Except for $\Delta \Phi=0$, no other notable quantitative differences are present between the cases $\Rey<1$ and $\Rey=20$. The general behaviour of both the antipleptic and symplectic MCW remains similar for all the other phase lags $\Delta \Phi$. Especially, the presence of the peak around $\Delta \Phi=\upi/4$ is present for all the Reynolds numbers tested ($\Rey=0.047$, $2$, $5$, $10$ not shown). The similarity of the metachronal cases (i.e. $\Delta\Phi \neq 0$) for this range of Reynolds numbers is certainly due to the fact that there are always cilia in the stroke phase. Hence, a transport is observed during all the beat cycle, and the inertial effects thus remain minor. This is no longer the case for fully synchronously beating cilia: despite being weak, inertial effects cancel the reversal of the flow.

\section{Conclusions}\label{sec:Concl}

By using a lattice Boltzmann solver coupled to an Immersed Boundary method, and considering a purely hydrodynamical feedback from the fluid, symplectic and antipleptic metachronal waves can emerge in a 3D two-phase flow configuration with a viscosity ratio of 20. It is known that metachronal waves may emerge due to hydrodynamic interactions \citep{Elgeti}. However, for the first time, both types of metachrony are observed to emerge in a two-layer fluid with different viscosities, using a simple feedback law. This feedback depends on a coupling parameter $\alpha$ that can be used to tune the strength and direction of the waves. It is observed that cilia experience weaker torques when being in a clustered configuration, i.e. during the recovery phase of antipleptic motion, as well as during the stroke phase of symplectic motion. 
The resulting beating pattern is a slow stroke phase and a fast recovery phase for both cases, suggesting that even if a simple hydrodynamic interaction is enough to let emerge metachronal waves, it is not sufficient to reproduce all features of the beating patterns observed in real ciliated surfaces, and other biological issues may play a role in the transport mechanism. The study of \citet{Hussong} has shown that metachronal motion can switch from symplectic to antipleptic with increasing inertial effects. Here, the influence of cilia density is highlighted: by assuming a least effort behaviour for the cilia, the present model shows that the metachrony can switch from antipleptic to symplectic by lowering the cilia density. A more detailed study of the quality of synchronization along the $x$ and $y$-directions is the next step toward a better understanding of the emergence of MCW.

A thorough comparative study of the antipleptic and symplectic MCW has been performed and the results show that the antipleptic MCW are the most efficient ones for transporting mucus. This is in accordance with most recent studies addressing this point \citep{Khaderi11, Osterman, Ding}; and now confirmed in a two-layer environment. In the range of cilia spacing studied, and especially for small phase lags ($|\Delta \Phi|<\upi/2$) the antipleptic MCW have a better ability to (i) transport the flow in the same direction compared to symplectic MCW with the same wavelength, (ii) generate higher flow rate, (iii) advect particles at a given power input, and (iv) generate a higher stretching and hence, are more able to mix fluids. On the contrary, symplectic MCW do not appear to have a great impact on the flow and are often less efficient than antipleptic MCW. This is not in agreement with \citet{Elgeti} who reported that symplectic waves are almost nearly as efficient as antipleptic waves, using an optimized-efficiency model. This difference may be due to several factors, including the two-layer flow character of the present work, the ratios of viscosity considered, or even the cilia beat shape used. However, these results are in accordance with \citet{Khaderi11} who explained this better transport efficiency as being due to vortices obstructing the flow for symplectic motions.

Among the quantities introduced earlier, some are more decisive for the transport of particles. The most important ones are the net volume of mucus displaced, the transport efficiency $\eta_{1}$, and the mixing capacity of the system. It has been shown all along this work that antipleptic MCW with $\Delta \Phi=\upi/4$ generate the larger flow while in the meantime showing the higher value of the transport efficiency (the power spent $P^{*}$ for $\Delta \Phi=\upi/4$ being minimal). A preliminary study of the mixing also strongly indicates that they might also be the more suitable to mix fluids. The best transport efficiency of the antipleptic MCW is certainly due to the clusterized aspect of the cilia in recovery phase which minimizes their impact on the flow, while cilia in stroke phase are able to fully exert their pushing action. While being more able to advect particles than any other form of coordinated motions, antipleptic MCW with $\Delta\Phi<\upi/2$ also require less power. Note that the results presented all along concern a viscosity ratio $r_{\nu}$ of 20. \citet{ChatelinPOncet} has shown that mucus transport was maximized for viscosity ratios ranging from 10 to 20. With the present model, and within this interval of viscosity ratios, the same trends are always observed (results not shown), and the best transport capacities are always obtained for antipleptic MCW with $\Delta \Phi=\upi/4$. The transport properties of the waves also remain similar and do not vary significantly. However, it is worth noticing that the best transport capacity was obtained for $r_{\nu}=20$. Hence, the optimum is probably outside the range of viscosity ratios studied. It also strongly indicates that increasing the mucus viscosity results in an increase of the transport properties. As in previous studies \citep{Norton, ChatelinPOncet}, an optimum viscosity for the transport of particles can certainly be found. A more detailed study of how different mucus viscosity affect the transport properties of the MCW is the next step toward a better understanding of the mucociliary clearance.

While in the present study only one prescribed beating pattern was investigated, the results can certainly be generalised to other beating patterns. An interesting path of investigation would be to study how this beat shape is influenced by different mucus viscosities, mucus rheologies, or thicknesses of the PCL. Note that a Reynolds number of 20 was used for computational reasons. While this does not affect the behaviour of both the antipleptic and symplectic MCW, weak inertial effects are seen to change the dynamics of the flow in the particular case of synchronously beating cilia.
Finally, the mucus was considered as being a Newtonian fluid in the present work, while it is in reality highly non-Newtonian. It is expected that the clearance velocity of both the antipleptic and symplectic MCW would be increased by considering the mucus as being a visco-elastic fluid. One could also expect the symplectic waves to become more efficient than the antipleptic MCW for highly viscous mucus. Indeed, the clusterized aspect of the cilia during the stroke phase of symplectic motion could help them overcome more easily the viscous resistance of the mucus \citep{Knight-Jones}. Nevertheless, note that studies \citep{Norton, ChatelinPOncet} all tend to show, however, that the mucus can not be too ``solid'' nor too ``liquid''.

The numerical results obtained here constitute a useful basis of investigation to progress in the understanding of respiratory diseases linked to cilia beating disorders, such as COPD or severe asthma \citep{chanez2005}. It is also of interest for industrial purposes, such as the design of cilia-based actuators for mixing \citep{Chen} or flow-regulator in microscopic biosensors.

Future works include the implementation of a realistic rheological model \citep{Lafforgue} for the mucus in order to take into account its highly non-Newtonian behaviour, and the study of the mass transfer occurring at the epithelial surface.
A long term objective of this work is to build a numerical environment to predict the transport and mixing of drugs inside both the mucus and PCL layers. With this tool, the effect of drugs could be tested virtually before proceeding to clinical trials. For instance, drugs acting on microscopic parameters such as beating frequency, viscosity of the mucus, density of cilia could be tested and their effects on macroscopic quantities such as mucus flow rate and mixing could be explored, in order to progress in the understanding and treatment of respiratory diseases.

The authors would like to thank Dr. Zhe Li and master intern Jean Mercat for the development of the numerical code. S. Chateau and S. Poncet acknowledge also the Natural Sciences and Engineering Research Council of Canada for its financial support through a Discovery Grant (RGPIN-2015-06512). This work was granted access to the HPC resources of Compute Canada and Aix-Marseille University (project Equip@Meso ANR-10-EQPX-29-01). 

%

\bibliographystyle{jfm}

\bibliography{jfm-instructions}

\begin{thebibliography}{57}
\expandafter\ifx\csname natexlab\endcsname\relax\def\natexlab#1{#1}\fi
\def\au#1{#1} \def\ed#1{#1} \def\yr#1{#1}\def\at#1{#1}\def\jt#1{\textit{#1}}
  \def\bt#1{#1}\def\bvol#1{\textbf{#1}} \def\vol#1{#1} \def\pg#1{#1}
  \def\publ#1{#1}\def\arxiv#1{#1}\def\org#1{#1}\def\st#1{\textit{#1}}

\bibitem[Blake(1971{\natexlab{{\em a\/}}})]{BlakeA}
{\sc \au{Blake, J.~R.}} \yr{1971{\natexlab{{\em a\/}}}}  \at{Infinite model for
  ciliary propulsion}.  \jt{J.~Fluid Mech.}  \bvol{49}~(2),  \pg{209--222}.

\bibitem[Blake(1971{\natexlab{{\em b\/}}})]{BlakeB}
{\sc \au{Blake, J.~R.}} \yr{1971{\natexlab{{\em b\/}}}}  \at{A spherical
  envelope approach to ciliary propulsion}.  \jt{J.~Fluid Mech.}
  \bvol{46}~(1),  \pg{199--208}.

\bibitem[Blake(1972)]{BlakeC}
{\sc \au{Blake, J.~R.}} \yr{1972}  \at{A model for the micro-structure in
  ciliated organisms}.  \jt{J.~Fluid Mech.}  \bvol{55}~(1),  \pg{1--23}.

\bibitem[Blake \& Chwang(1974)]{BlakeD}
{\sc \au{Blake, J.~R.} \& \au{Chwang, A.~T.}} \yr{1974}  \at{Fundamental
  singularities of viscous flow. \mbox{P}art \mbox{II}}.  \jt{J.~Eng. Math.}
  \bvol{8}~(1),  \pg{113--124}.

\bibitem[Brennen \& Winet(1977)]{brennen}
{\sc \au{Brennen, C.} \& \au{Winet, H.}} \yr{1977}  \at{Fluid mechanics of
  propulsion by cilia and flagella}.  \jt{Annu.~Rev. Fluid Mech.}  \bvol{9},
  \pg{339--398}.

\bibitem[Chanez(2005)]{chanez2005}
{\sc \au{Chanez, P.}} \yr{2005}  \at{Severe asthma is an epithelial disease}.
  \jt{Eur.~Respir. J.}  \bvol{25}~(6),  \pg{945--946}.

\bibitem[Chatelin(2013)]{Chatelin}
{\sc \au{Chatelin, R.}} \yr{2013}  \at{M\'ethodes num\'eriques pour
  l'\'ecoulement de \mbox{S}tokes 3\mbox{D} : fluides \`a viscosit\'e variable
  en g\'eom\'etrie complexe mobile; \mbox{A}pplication aux fluides
  biologiques}. PhD thesis, Institut de Math\'ematiques de Toulouse.

\bibitem[Chatelin \& Poncet(2016)]{ChatelinPOncet}
{\sc \au{Chatelin, R.} \& \au{Poncet, P.}} \yr{2016}  \at{A parametric study of
  mucociliary transport by numerical simulations of 3\mbox{D} non-homogeneous
  mucus}.  \jt{J.~Biomech.}  \bvol{49}~(9),  \pg{1772--1780}.

\bibitem[Chen {\em et~al.\/}(2013)Chen, Chen, Lin \& Hu]{Chen}
{\sc \au{Chen, C.~Y.}, \au{Chen, C.~Y.}, \au{Lin, C.~Y.} \& \au{Hu, Y.~T.}}
  \yr{2013}  \at{Magnetically actuated artificial cilia for optimum mixing
  performance in microfluidics}.  \jt{Lab Chip}  \bvol{13}~(14),
  \pg{2834--2839}.

\bibitem[Dauptain {\em et~al.\/}(2008)Dauptain, Favier \& Bottaro]{Dauptain}
{\sc \au{Dauptain, A.}, \au{Favier, J.} \& \au{Bottaro, A.}} \yr{2008}
  \at{Hydrodynamics of ciliary propulsion}.  \jt{J.~Fluid. Struct.}
  \bvol{24}~(8),  \pg{1156--1165}.

\bibitem[Ding {\em et~al.\/}(2014)Ding, Nawroth, McFall-Ngai \& Kanso]{Ding}
{\sc \au{Ding, Y.}, \au{Nawroth, J.~C.}, \au{McFall-Ngai, M.~J.} \& \au{Kanso,
  E.}} \yr{2014}  \at{Mixing and transport by ciliary carpets: a numerical
  study}.  \jt{J.~Fluid Mech.}  \bvol{743},  \pg{124--140}.

\bibitem[Downton \& Stark(2009)]{Downton}
{\sc \au{Downton, M.~T.} \& \au{Stark, H.}} \yr{2009}  \at{Beating kinematics
  of magnetically actuated cilia}.  \jt{Europhys. Lett.}  \bvol{85}~(44002).

\bibitem[Elgeti \& Gompper(2013)]{Elgeti}
{\sc \au{Elgeti, J.} \& \au{Gompper, G.}} \yr{2013}  \at{Emergence of
  metachronal waves in cilia arrays}.  \jt{PNAS}  \bvol{110}~(12),
  \pg{4470--4475}.

\bibitem[Eloy \& Lauga(2012)]{Eloy}
{\sc \au{Eloy, C.} \& \au{Lauga, E.}} \yr{2012}  \at{Kinematics of the most
  efficient cilium}.  \jt{Phys.~Rev. Lett.}  \bvol{109}~(038101).

\bibitem[Fauci \& Dillon(2006)]{Fauci}
{\sc \au{Fauci, L.~J.} \& \au{Dillon, R.}} \yr{2006}  \at{Biofluidmechanics of
  reproduction}.  \jt{Annu.~Rev. Fluid Mech.}  \bvol{38},  \pg{371--394}.

\bibitem[Gardiner(2005)]{Gardiner}
{\sc \au{Gardiner, M.~B.}} \yr{2005}  \at{The importance of being cilia}.
  \jt{HHMI Bulletin}  \bvol{64},  \pg{32--36}.

\bibitem[Gauger {\em et~al.\/}(2009)Gauger, Downton \& Stark]{Gauger2009}
{\sc \au{Gauger, E.~M.}, \au{Downton, M.~T.} \& \au{Stark, H.}} \yr{2009}
  \at{Fluid transport at low \mbox{R}eynolds number with magnetically actuated
  artificial cilia}.  \jt{Eur.~Phys. J. E}  \bvol{28}~(2),  \pg{231--242}.

\bibitem[Gueron \& Levit-Gurevich(1999)]{Gueron99}
{\sc \au{Gueron, S.} \& \au{Levit-Gurevich, K.}} \yr{1999}  \at{Energetic
  considerations of ciliary beating and the advantage of metachronal
  coordination}.  \jt{PNAS}  \bvol{96}~(22),  \pg{12240--12245}.

\bibitem[Gueron {\em et~al.\/}(1997)Gueron, Levit-Gurevich, Liron \&
  Blum]{Gueron97}
{\sc \au{Gueron, S.}, \au{Levit-Gurevich, K.}, \au{Liron, N.} \& \au{Blum,
  J.~J.}} \yr{1997}  \at{Cilia internal mechanism and metachronal coordination
  as the result of hydrodynamical coupling}.  \jt{PNAS}  \bvol{94}~(12),
  \pg{6001--6006}.

\bibitem[Guo {\em et~al.\/}(2014)Guo, Nawroth, Ding \& Kanso]{GuoH}
{\sc \au{Guo, H.}, \au{Nawroth, J.}, \au{Ding, Y.} \& \au{Kanso, E.}} \yr{2014}
   \at{Cilia beating patterns are not hydrodynamically optimal}.  \jt{Phys.~
  Fluids}  \bvol{26}~(091901).

\bibitem[Guo \& Shu(2013)]{GuoZHAOLI}
{\sc \au{Guo, Z.} \& \au{Shu, C.}} \yr{2013} {\em Lattice Boltzmann Method and
  Its Applications in Engineering\/}.  \publ{World Scientific Publishing
  Company, Singapore}.

\bibitem[Hussong {\em et~al.\/}(2011)Hussong, Breugem \& Westerweel]{Hussong}
{\sc \au{Hussong, J.}, \au{Breugem, W.P.} \& \au{Westerweel, J.}} \yr{2011}
  \at{A continuum model for flow induced by metachronal coordination between
  beating cilia}.  \jt{J.~Fluid Mech.}  \bvol{684},  \pg{137--162}.

\bibitem[Keller \& Brennen(1968)]{Keller}
{\sc \au{Keller, S.~R.} \& \au{Brennen, C.}} \yr{1968} {\em A traction-layer
  model for ciliary propulsion\/}.  \publ{Proceedings of the Symposium on
  Swimming and Flying in Nature, held at the \mbox{C}alifornia \mbox{I}nstitute
  of \mbox{T}echnology, \mbox{P}asadena, 253--271, Plenum Press, New York}.

\bibitem[Kelley \& Ouellette(2011)]{Kelley}
{\sc \au{Kelley, D.~H.} \& \au{Ouellette, N.~T.}} \yr{2011}  \at{Separating
  stretching from folding in fluid mixing}.  \jt{Nat.~Phys.}  \bvol{7},
  \pg{477--480}.

\bibitem[Khaderi {\em et~al.\/}(2010)Khaderi, Baltussen, Anderson, den Toonder
  \& Onck]{KhaderiBalt}
{\sc \au{Khaderi, S.~N.}, \au{Baltussen, M. G. H.~M.}, \au{Anderson, P.~D.},
  \au{den Toonder, J. M.~J.} \& \au{Onck, P.~R.}} \yr{2010}  \at{Breaking of
  symmetry in microfluidic propulsion driven by artificial cilia}.  \jt{Phys.
  Rev. E}  \bvol{82}~(027302).

\bibitem[Khaderi {\em et~al.\/}(2011)Khaderi, Den-Toonder \& Onck]{Khaderi11}
{\sc \au{Khaderi, S.~N.}, \au{Den-Toonder, J. M.~J.} \& \au{Onck, P.~R.}}
  \yr{2011}  \at{Microfluidic propulsion by the metachronal beating of magnetic
  artificial cilia: a numerical analysis}.  \jt{J.~Fluid Mech.}  \bvol{688},
  \pg{44--65}.

\bibitem[Kim \& Netz(2006)]{KimNetz}
{\sc \au{Kim, Y.~W.} \& \au{Netz, R.~R.}} \yr{2006}  \at{Pumping fluids with
  periodically beating grafted elastic filaments}.  \jt{Phys. Rev. Lett.}
  \bvol{96}~(15),  \pg{158101}.

\bibitem[Kirkham {\em et~al.\/}(2002)Kirkham, Sheehan, Knight, Richardson \&
  Thornton]{Kirkham}
{\sc \au{Kirkham, S.}, \au{Sheehan, J.~K.}, \au{Knight, D.}, \au{Richardson,
  P.~S.} \& \au{Thornton, D.~J.}} \yr{2002}  \at{Heterogeneity of airways
  mucus: variations in the amounts and glycoforms of the major oligomeric
  mucins \mbox{MUC5AC} and \mbox{MUC5B}}.  \jt{Biochem.~J.}  \bvol{361}~(3),
  \pg{537--546}.

\bibitem[Knight-Jones(1954)]{Knight-Jones}
{\sc \au{Knight-Jones, E.W.}} \yr{1954}  \at{Relations between metachronism and
  the direction of ciliary beat in \mbox{M}etazoa}.  \jt{J.~Cell Sci.}
  \bvol{s3--95},  \pg{503--521}.

\bibitem[Lafforgue {\em et~al.\/}(2016)Lafforgue, Poncet, Seyssiecq-Guarente \&
  Favier]{Lafforgue}
{\sc \au{Lafforgue, O.}, \au{Poncet, S.}, \au{Seyssiecq-Guarente, I.} \&
  \au{Favier, J.}} \yr{2016} {\em Rheological characterization of
  macromolecular colloidal gels as simulant of bronchial mucus\/}.
  \publ{32$^{nd}$ International Conference of the Polymer Processing Society
  (PPS-32), Lyon, France}.

\bibitem[Lai {\em et~al.\/}(2009)Lai, Wang, Wirtz \& Hanes]{Lai}
{\sc \au{Lai, S.~K.}, \au{Wang, Y.~Y.}, \au{Wirtz, D.} \& \au{Hanes, J.}}
  \yr{2009}  \at{Micro- and macrorheology of mucus}.  \jt{Adv.~Drug Deliver
  Rev.}  \bvol{61}~(2),  \pg{86--100}.

\bibitem[Lauga \& Eloy(2013)]{Lauga}
{\sc \au{Lauga, E.} \& \au{Eloy, C.}} \yr{2013}  \at{Shape of optimal active
  flagella}.  \jt{J.~Fluid Mech.}  \bvol{730}~(R1),  \pg{1--11}.

\bibitem[Li {\em et~al.\/}(2009)Li, Tan \& Zhang]{Huaming}
{\sc \au{Li, H.}, \au{Tan, J.} \& \au{Zhang, M.}} \yr{2009}  \at{Dynamics
  modeling and analysis of a swimming microrobot for controlled drug delivery}.
   \jt{IEEE T-ASE}  \bvol{6}~(2),  \pg{220--227}.

\bibitem[Li {\em et~al.\/}(2016)Li, Favier, D'Ortona \& Poncet]{JCPZHE}
{\sc \au{Li, Z.}, \au{Favier, J.}, \au{D'Ortona, U.} \& \au{Poncet, S.}}
  \yr{2016}  \at{An improved explicit immersed boundary method to couple with
  lattice \mbox{B}oltzmann model for single- and multi-component fluid flows}.
  \jt{J.~Comput. Phys.}  \bvol{304},  \pg{424--440}.

\bibitem[Lighthill(1976)]{Lighthill}
{\sc \au{Lighthill, J.}} \yr{1976}  \at{Flagellar hydrodynamics}.  \jt{SIAM
  Rev.}  \bvol{18},  \pg{161--230}.

\bibitem[Lukens {\em et~al.\/}(2010)Lukens, Yang \& Fauci]{Lukens}
{\sc \au{Lukens, S.}, \au{Yang, X.} \& \au{Fauci, L.}} \yr{2010}  \at{Using
  lagrangian coherent structures to analyze fluid mixing by cilia}.  \jt{Chaos}
   \bvol{20}~(1),  \pg{017511}.

\bibitem[Matsui {\em et~al.\/}(1998)Matsui, Randell, Peretti, Davis \&
  Boucher]{Matsui}
{\sc \au{Matsui, H.}, \au{Randell, S.~H.}, \au{Peretti, S.~W.}, \au{Davis,
  W.~C.} \& \au{Boucher, R.~C.}} \yr{1998}  \at{Coordinated clearance of
  periciliary liquid and mucus from airway surfaces}.  \jt{J.~Clin. Invest.}
  \bvol{102}~(6),  \pg{1125--1131}.

\bibitem[Mitran(2007)]{Mitran}
{\sc \au{Mitran, S.M.}} \yr{2007}  \at{Metachronal wave formation in a model of
  pulmonary cilia}.  \jt{Comput.~Struct.}  \bvol{85}~(11-14),  \pg{763--774}.

\bibitem[Niedermayer {\em et~al.\/}(2008)Niedermayer, Eckhardt \&
  Lenz]{Niedermayer}
{\sc \au{Niedermayer, T.}, \au{Eckhardt, B.} \& \au{Lenz, P.}} \yr{2008}
  \at{Synchronization, phase locking, and metachronal wave formation in ciliary
  chains}.  \jt{Chaos}  \bvol{18}~(3),  \pg{037128}.

\bibitem[Norton {\em et~al.\/}(2011)Norton, Robinson \& Weinstein]{Norton}
{\sc \au{Norton, M.~M.}, \au{Robinson, R.~J.} \& \au{Weinstein, S.~J.}}
  \yr{2011}  \at{Model of ciliary clearance and the role of mucus rheology}.
  \jt{Phys.~Rev. E}  \bvol{83}~(011921).

\bibitem[Osterman \& Vilfan(2011)]{Osterman}
{\sc \au{Osterman, N.} \& \au{Vilfan, A.}} \yr{2011}  \at{Finding the ciliary
  beating pattern with optimal efficiency}.  \jt{PNAS}  \bvol{108}~(38),
  \pg{15727--15732}.

\bibitem[Phan-Thien {\em et~al.\/}(1987)Phan-Thien, Tran-Cong \& Ramia]{Phan}
{\sc \au{Phan-Thien, N.}, \au{Tran-Cong, T.} \& \au{Ramia, M.}} \yr{1987}
  \at{A boundary-element analysis of flagellar propulsion}.  \jt{J.~Fluid
  Mech.}  \bvol{184},  \pg{533--549}.

\bibitem[Porter {\em et~al.\/}(2012)Porter, Coon, Kang, Moulton \&
  Carey]{Porter}
{\sc \au{Porter, M.~L.}, \au{Coon, E.~T.}, \au{Kang, Q.}, \au{Moulton, J.~D.}
  \& \au{Carey, J.~W.}} \yr{2012}  \at{Multicomponent interparticle-potential
  lattice \mbox{B}oltzmann model for fluids with large viscosity ratios}.
  \jt{Phys. Rev. E}  \bvol{86}~(036701).

\bibitem[Reynolds(1965)]{Reynolds}
{\sc \au{Reynolds, A.~J.}} \yr{1965}  \at{The swimming of minute organisms}.
  \jt{J.~Fluid Mech.}  \bvol{23}~(2),  \pg{241--260}.

\bibitem[Sanderson \& Sleigh(1981)]{Sanderson}
{\sc \au{Sanderson, M.J.} \& \au{Sleigh, M.A.}} \yr{1981}  \at{Ciliary activity
  of cultured rabbit tracheal epithelium: \mbox{B}eat pattern and metachrony}.
  \jt{J.~Cell. Sci.}  \bvol{47},  \pg{331--341}.

\bibitem[Satir \& Christensen(2007)]{Satir}
{\sc \au{Satir, P.} \& \au{Christensen, S.}} \yr{2007}  \at{Overview of
  structure and function of mammalian cilia}.  \jt{Annu.~Rev. Physiol.}
  \bvol{69},  \pg{377--400}.

\bibitem[Sedaghat {\em et~al.\/}(2016)Sedaghat, Shahmardan, Norouzi,
  Jayathilake \& Nazari]{Sedaghat}
{\sc \au{Sedaghat, M.~H.}, \au{Shahmardan, M.~M.}, \au{Norouzi, M.},
  \au{Jayathilake, P.~G.} \& \au{Nazari, M.}} \yr{2016}  \at{Numerical
  simulation of muco-ciliary clearance: immersed boundary lattice
  \mbox{B}oltzmann method}.  \jt{Comput.~Fluids}  \bvol{131},  \pg{91--101}.

\bibitem[Shan \& Chen(1993)]{Shan1993}
{\sc \au{Shan, X.} \& \au{Chen, H.}} \yr{1993}  \at{Lattice boltzmann model for
  simulating flows with multiple phases and components}.  \jt{Phys.~Rev. E}
  \bvol{47}~(3),  \pg{1815--1819}.

\bibitem[Shan \& Chen(1994)]{ShanChen}
{\sc \au{Shan, X.} \& \au{Chen, H.}} \yr{1994}  \at{Simulation of nonideal
  gases and liquid-gas phase transitions by the lattice \mbox{B}oltzmann
  equation}.  \jt{Phys.~Rev. E}  \bvol{49}~(4),  \pg{2941--2948}.

\bibitem[Sleigh(1962)]{Sleigh}
{\sc \au{Sleigh, M.~A.}} \yr{1962} {\em The biology of Cilia and Flagella\/}.
  \publ{Pergamon Press, Oxford}.

\bibitem[Sleigh {\em et~al.\/}(1988)Sleigh, Blake \& Liron]{Sleigh2}
{\sc \au{Sleigh, M.~A.}, \au{Blake, J.~R.} \& \au{Liron, N.}} \yr{1988}
  \at{The propulsion of mucus by cilia}.  \jt{Am.~Rev. Respir. Dis.}
  \bvol{137}~(3),  \pg{726--741}.

\bibitem[Smith {\em et~al.\/}(2007)Smith, Gaffney \& Blake]{Smith}
{\sc \au{Smith, D.~J.}, \au{Gaffney, E.~A.} \& \au{Blake, J.~R.}} \yr{2007}
  \at{Discrete cilia modelling with singularity distributions: Application to
  the embryonic node and the airway surface liquid}.  \jt{B.~Math. Biol.}
  \bvol{69}~(5),  \pg{1477--1510}.

\bibitem[Taylor(1951)]{Taylor}
{\sc \au{Taylor, G.}} \yr{1951}  \at{Analysis of the swimming of microscopic
  organisms}.  \jt{Proc.~R. Soc. A}  \bvol{209}~(1099),  \pg{447--461}.

\bibitem[Tuck(1968)]{Tuck}
{\sc \au{Tuck, E.~O.}} \yr{1968}  \at{A note on a swimming problem}.
  \jt{J.~Fluid Mech.}  \bvol{31}~(2),  \pg{305--308}.

\bibitem[Widdicombe \& Widdicombe(1995)]{Widi}
{\sc \au{Widdicombe, J.~H.} \& \au{Widdicombe, J.~G.}} \yr{1995}
  \at{Regulation of human airway surface liquid}.  \jt{Resp.~ Physiol.}
  \bvol{99}~(1),  \pg{3--12}.

\bibitem[Winters \& Yeates(1997)]{Winters}
{\sc \au{Winters, S.~L.} \& \au{Yeates, D.~B.}} \yr{1997}  \at{Roles of
  hydration, sodium, and chloride in regulation of canine mucociliary transport
  system}.  \jt{J.~Appl. Physiol.}  \bvol{83}~(4),  \pg{1360--1369}.

\bibitem[Zou {\em et~al.\/}(1995)Zou, Hou, Chen \& Doolen]{Zou}
{\sc \au{Zou, Q.}, \au{Hou, S.}, \au{Chen, S.} \& \au{Doolen, G.~D.}} \yr{1995}
   \at{A improved incompressible lattice \mbox{B}oltzmann model for
  time-independent flows}.  \jt{J.~Stat. Phys.}  \bvol{81}~(1),  \pg{35--48}.

\end{thebibliography}

\end{document}